\documentclass[pdflatex,sn-mathphys-num]{sn-jnl}
\usepackage{graphicx}%
\usepackage{multirow}%
\usepackage{amsmath,amssymb,amsfonts}%
\usepackage{amsthm}%
\usepackage{mathrsfs}%
\usepackage[title]{appendix}%
\usepackage{xcolor}%
\usepackage{textcomp}%
\usepackage{manyfoot}%
\usepackage{booktabs}%
\usepackage{algorithm}%
\usepackage{algorithmicx}%
\usepackage{algpseudocode}%
\usepackage{listings}%
\usepackage{longtable}%
\usepackage{calc}
\usepackage{array}
\usepackage{pdfpages}

\begin{document} 

\title[Article Title]{Pulse-by-pulse imaging of reversible, collective reshaping in
laser-excited gold nanorods} 
\author[1]{\fnm{Mees} \sur{Dieperink}}  \affil[1]{\orgdiv{Center for Nanophotonics}, \orgname{NWO-Institute AMOLF}, \orgaddress{\city{Amsterdam}, \country{Netherlands}}} 
\author[1]{\fnm{Loriane} \sur{Monin}} 
\author[2]{\fnm{Ichiro} \sur{Ohnishi}} \affil[2]{\orgdiv{EM Business Unit}, \orgname{JEOL}, \orgaddress{\city{Tokyo}, \country{Japan}}} 
\author[2]{\fnm{Takeo} \sur{Sasaki}} 
\author[2]{\fnm{Yu} \sur{Jimbo}} 
\author[1]{\fnm{Erik} \sur{Garnett}} 
\author*[1]{\fnm{Wiebke} \sur{Albrecht}} \email{albrecht@amolf.nl}

\abstract{ Nanomaterial properties are governed by atomic structure, and driving structural changes on demand could enable materials with programmable properties. Pulsed light is the most controllable route: wavelength, fluence, polarization and pulse timing can each be tuned to inject energy with nanometre and picosecond precision and mould atomic positions. Yet the spatiotemporal dynamics of these transformations have remained hidden, inferred only from before-and-after snapshots. Here we track gold nanorods in a transmission electron microscope under pulsed optical excitation, quantifying the structural transition frame by frame. The rods enter an unexpectedly dynamic regime: material redistributes within and between particles, twin boundaries nucleate and propagate, rods hundreds of nanometres apart reshape in concert, and individual rods almost vanish before recovering their original shape, indicating a structural memory. Driving and following these transformations with pulse-synchronized in situ electron microscopy could turn light-induced reshaping into a tool for setting structure, and consequently properties, on demand.} 


\maketitle 
\newpage
\section{Introduction}\label{introduction}

At the nanoscale, the equilibrium rigidity of bulk crystals gives way to
a landscape of metastable configurations whose energetics are dominated
by surfaces, interfaces, and defects. The large surface-to-volume ratio
of nanocrystals lowers the cost of rearranging atoms, so that kinetics
can outweigh thermodynamics in setting the accessible structural states.
This sensitivity is central to catalysis \cite{dubau2016defects,jia2019role}, plasmonics, optics,
magnetism \cite{lak2021embracing,oliveira20203d}, and mechanical response, where small changes in surface
structure or defect density produce large functional shifts. Actively
manipulating atomic configurations therefore offers a route to materials
whose functional properties can be switched or tuned in real time.
Dynamic and resonant catalysis is one of the most prominent application
areas that stands to benefit: by oscillating the structure, strain, or
electronic state of a catalyst at the timescale of turnover, rates and
selectivities inaccessible at steady state become reachable \cite{monai2026grand}. Structural
modulation is a particularly powerful lever, able to dynamically optimize binding energies for sequential steps within a single catalytic
cycle \cite{wittreich2022catalytic}.

Pulsed light is an attractive route to such modulation, particularly in
metal nanoparticles, where localized surface plasmon resonances (LSPRs)
enable strong, tunable coupling between light, electrons, and phonons. Repetition rate, pulse duration, and fluence can be independently controlled, and absorption of pulses generates steep, confined
temperature and stress transients\cite{staechelin2021size,chiang2023electron}, driving rapid heating-cooling
cycles that quench nanoparticles into metastable configurations
inaccessible through conventional synthesis \cite{gonzalez2022engineering}. Pulsed illumination has
thus been used to form hollow structures, redistribute alloy components,
introduce twin boundaries and stacking faults, and transform
facets \cite{albrecht20213d, gonzalez2021controlled, gonzalez2017femtosecond, manzaneda2023multi, milagres20203d, gonzalez2020formation, gonzalez2016reshaping, gonzalez2015femtosecond, castro2020hollow, diaz2018using, link2000does, yan2020light, sumimoto2014situ}. Yet the pathways underlying these transformations remain
largely inferred from before-and-after snapshots or simulations \cite{opletal2011stability,wang2005surface},
and techniques to measure directly how atoms move, defects form, and
shapes evolve during excitation, and whether these are controllable or
reversible, are still scarce.

Transmission electron microscopy (TEM) is uniquely suited to visualizing
these dynamics at the relevant scales. To date, insight has come largely
from static, correlative approaches: combining pulsed laser irradiation
with before-and-after electron tomography has shown that optical
excitation can drive gold nanoparticles into kinetically trapped 3D
atomic configurations inaccessible through thermal processing alone \cite{albrecht20213d},
establishing pulsed light as a genuine tool for programming
nanostructures. Nevertheless, the structural dynamics within each pulse,
how atoms move, defects propagate, and configurations are selected, have
so far remained hidden. The emergence of \emph{in situ} TEM under light
excitation now makes it possible to follow these processes in real time,
opening the door from observing the outcomes of pulsed excitation to
actively controlling them \cite{zak2022light}.

Here, by combining pulsed optical excitation with a pulse-synchronized TEM acquisition scheme, we capture the structural state of gold nanorods after every individual optical pulse and develop a framework to quantify the resulting reshaping trajectories. We find an unexpectedly dynamic regime of correlated, partly reversible restructuring spanning single particles to hundreds of nanometres, including a structural memory in which specific configurations are repeatedly recovered. These results establish pulse-synchronized in situ TEM as a platform for resolving non-equilibrium structural dynamics one excitation event at a time, opening a path toward actively steering nanoscale structure, and thereby function, through pulsed optical excitation.

\section{Main}

\subsection{Pulse-synchronized optical excitation and global
evolution}\label{pulse-synchronized-optical-excitation-and-global-evolution}

\begin{figure}[H]
\centering
\includegraphics[width=0.8\textwidth]{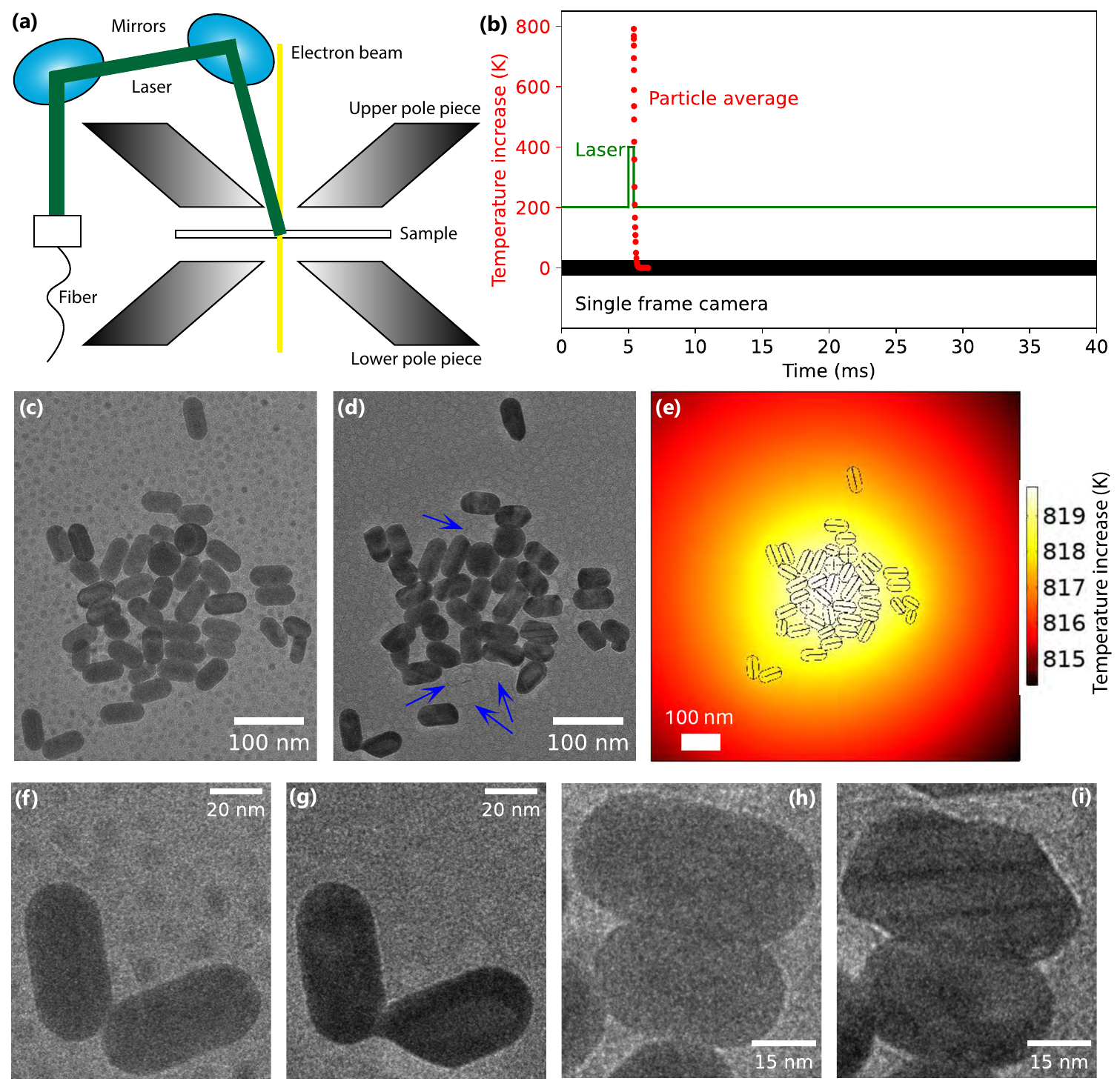}
\caption{Experimental platform and before -- after
response to pulsed optical excitation. (a) A schematic drawing of
light-incoupling into a TEM column. (b) Schematic of the
pulse-synchronized TEM imaging scheme: each 400\,$\mu$s laser pulse is
embedded in a 40\,ms camera frame, enabling single-pulse structural
snapshots. The laser pulse length is indicated in green. The resulting
temperature increase and decay in time that the particles consequently
experience is plotted in red (see Fig. S4 for zoom in). (c) Last
stable frame before the start of the optical excitation, showing the
initial morphology of the nanorod cluster. (d) Final frame after
prolonged pulsed illumination, demonstrating that rods reshaped,
disappeared from projection (blue arrows), or developed twinning
defects. (e) Peak temperature obtained from COMSOL heat-diffusion
simulations under the experimental pulse conditions. Temperatures
increase by 820\,K up to \textasciitilde1115\,K, sufficient to enable
rapid surface diffusion and partial melting in nanoscale gold. (f) Zoom
into the two nanorods at the bottom left of the field of view in (c)
before optical excitation. (g) The same two nanorods after the
experiment exemplifying structural reshaping. (h) Zoom into two nanorods
at the edge of the bigger cluster in (c) before optical excitation. (g)
The same two nanorods after the experiment highlighting the development
of twinning defects.}\label{fig1}
\end{figure}

Gold nanorods on amorphous carbon TEM grids were excited \emph{in situ}
inside a TEM with a laser-incoupling system (Fig. 1a, Methods). A
modulated CW laser tuned near the longitudinal LSPR (Fig. S1) delivered
400\,$\mu$s optical pulses synchronized to the 40\,ms camera frame, so that
each frame captures the structural state of exactly one pulse (green
line in Fig. 1b) resulting in a temperature spike (red dots in Fig. 1b).
Because the laser was active for only \textasciitilde1\% of each frame,
every image is a time-average dominated by the laser-off periods, during
which the nanorods are static, ensuring reshaping is captured free of
motion blur.

Figure 1c shows the region of interest immediately before illumination:
a compact ensemble of Au nanorods with homogeneous contrast within
individual rods, confirming their known single-crystalline fcc
character \cite{albrecht20213d}. The assembly spans a range of interparticle configurations,
from fully isolated rods to closely spaced pairs and dense clusters,
providing a natural laboratory for observing how proximity governs
reshaping dynamics. Variations in apparent intensity between rods
reflect differences in thickness and diffraction contrast: rods oriented
near a zone axis appear darker due to enhanced Bragg scattering, while
those tilted away appear brighter. Upon pulsed excitation for 76.44\,s
with 1911 pulses (1911 frames), the ensemble underwent extensive
structural rearrangement (Fig. 1d): several nanorods reshaped
substantially (zoom in example in Fig. 1f, g), some disappeared from
projection (blue arrows), and twinning defects emerged across multiple
particles seen as dark lines (zoom in example in Fig. 1h, i). A
differential contrast map of the overview images before and after
excitation (Fig. 1c,d, segmentation in Fig. S2 and difference in Fig. S3a) shows adjacent regions of
gain and loss across the ensemble, indicating redistribution of
material. Segmenting the particles and integrating this contrast gives
the change in projected area per particle (Fig. S3b): the particles that
grew do not offset those that shrank or disappeared, leaving a net loss
in projection consistent with some sublimation. Since both measures are
projections, they provide an estimate of redistributed material rather
than a volume balance.

To assess whether this reshaping is thermally plausible, we performed
steady-state heat-diffusion simulations (Fig. 1e). Although excitation
is near the longitudinal plasmon resonance, the dominant heat source is
photothermal absorption in the carbon support, with an additional but
smaller contribution from direct nanorod absorption. Together these
yield a peak temperature of \textasciitilde1115 K during each pulse,
below the bulk melting point of gold (1337\,K), but high enough for Au
atoms, particularly at the surface, to be highly mobile \cite{imura2025ultrafine, petrova2006temperature, cho2020atomic}. The 400\,$\mu$s heating window is orders of magnitude longer than the nanoscale
melting and rearrangement times of gold (ps-ns regime), providing an
extended high-temperature window for atomic diffusion and defect
reconfiguration within each pulse. Time-dependent simulations indicate
that the system returns to near room temperature within
\textasciitilde200\,$\mu$s (Fig. S4), well within the 40\,ms between pulses
and consistent with each pulse acting as a largely independent
excitation event. Both the peak temperature and the cooling time depend
on the optical absorption and thermal conductivity of the thin amorphous
carbon support, for which reported values span a wide range and which is
expected to graphitise progressively under pulsed illumination. We
therefore treat these temperatures as representative rather than exact,
and cannot fully exclude a small residual inter-pulse temperature rise.

\subsection{From twin nucleation to collective mass
transport}\label{from-twin-nucleation-to-collective-mass-transport}

\begin{figure}[H]
\centering
\includegraphics[width=0.8\textwidth]{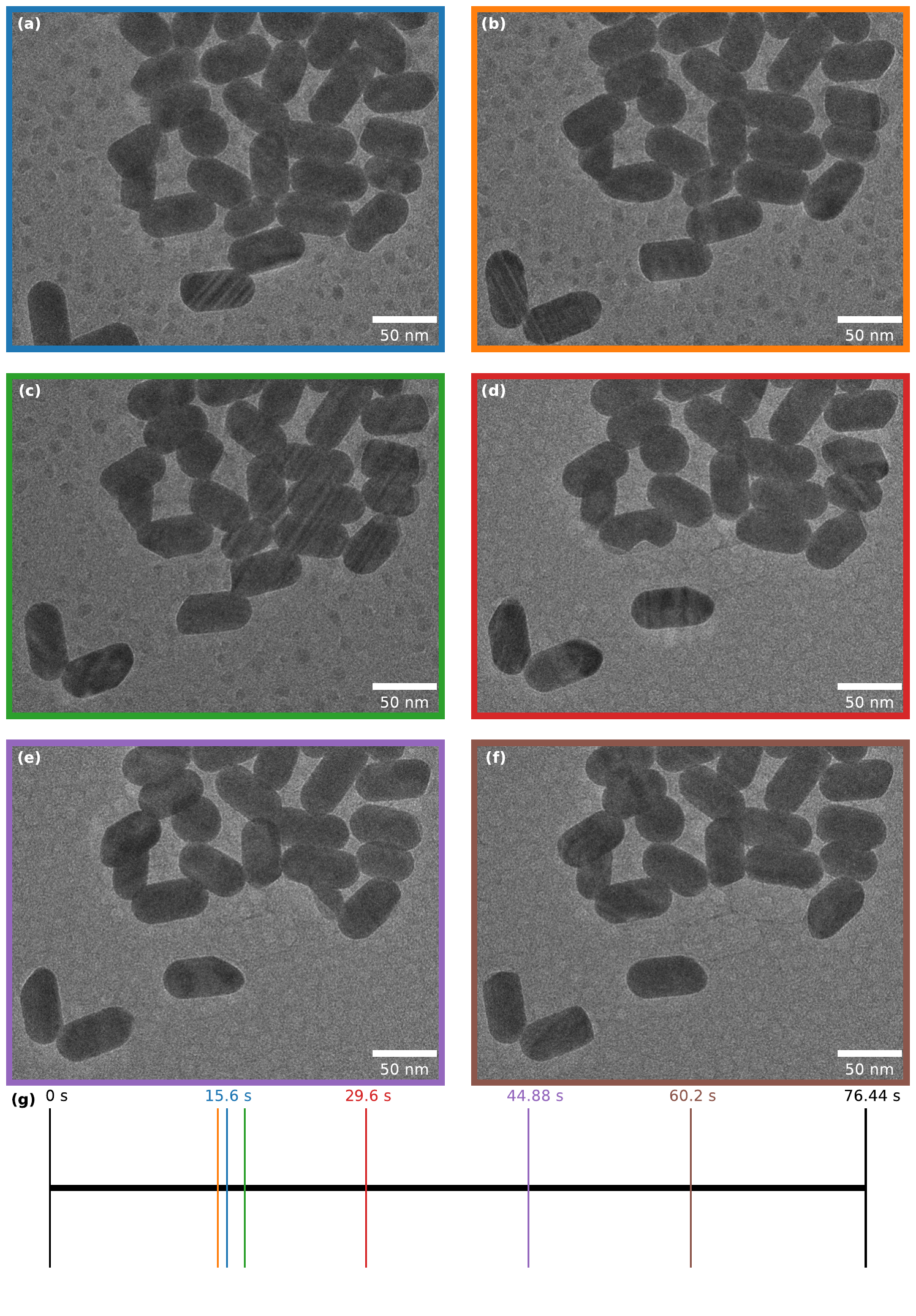}
\caption{Pulse-resolved evolution of twinning and
collective structural reorganization. (a) Early-stage twinning: a
single rod exhibits a distinct \{111\}-type twin boundary. (b)
Intermediate stage: twin boundaries appear in adjacent rods, indicating
localized crystallographic alignment. (c) Late stage: extended regions
of the cluster share similarly oriented twinning defects, demonstrating
large-scale structural reorientation consistent with highly mobile atoms
during pulsed excitation. (d) Particle deformation after twinning. (e,f
Particle disappearance. (g) Relative timing of frames during experiment
(colors correspond to a-f).}\label{fig2}
\end{figure}

Having established the global thermal picture, we next resolve how the
structure evolves pulse-to-pulse (Fig. 2, Supplementary Movie ). Within
the first few seconds, which is an upper bound due to heat-induced drift
of the support preventing registration of the earliest frames, the
cluster enters a mobile regime: individual rods begin to reshape and
grow visibly more dynamic from frame to frame. A prominent structural
signature is the emergence of crystallographic twin boundaries, which
develops from isolated rods into ensemble-wide correlation. Twinning
first appears sporadically in individual rods (Fig. 2a); later,
boundaries in adjacent rods align (Fig. 2b), and after \textasciitilde15
s a single boundary is traced continuously across several particles
(Fig. 2c). Coherent twin boundaries in fcc Au lie on \{111\} planes, so
continuity across initially differently oriented rods requires more than
contact: the gold in at least one particle must have reoriented,
abandoning its original crystallographic orientation. Together with the
redistribution evident from the vanishing rods and difference maps (Fig.
S3a), this points to a structural ratchet: each pulse merges
neighbouring rods a little further, so that gold can be transported over
increasing distances.

Consistent with this, the rearrangements extend over increasing
distances as the experiment proceeds: entire particles disappear without
a corresponding gain in their immediate neighbours (Fig. 2d-f), while
many rods across the cluster each grow slightly in projected area (Fig.
S3), indicating redistribution over long distances rather than local
transfer. The twin orientations also reset from one pulse to the next,
and do so in concert across connected particles (Fig. S5d,e), indicating
that the lattice remains highly mobile within each pulse and that merged
particles reorient as a single crystalline unit rather than
independently. Throughout, the outer envelope of the cluster remains
largely constrained, plausibly by a carbonaceous shell that forms on the
particle surfaces through beam-induced carbonization of the surface
ligands and contamination, leaving well-defined internal shapes and
channels along which atoms can redistribute \cite{albrecht2018impact}.

\subsection{Pulse-resolved reshaping of an isolated
nanorod}\label{pulse-resolved-reshaping-of-an-isolated-nanorod}

The collective dynamics above call for a quantitative, frame-by-frame
analysis with a metric robust to segmentation noise. We begin with the
simplest case, an isolated rod, where reshaping can be tracked without
mass exchange with neighbours. For NR 6 (as labelled in Fig. S6), an
isolated rod that stays highly active throughout, we split its
segmentation mask into two tip halves and track the projected area
retained in each; their difference defines a tip-asymmetry signal that
is near zero for symmetric tips and deviates when one tip gains or loses
area (SI). To separate genuine reshaping from measurement noise, we
calibrate a noise band from the residual fluctuations of the static
reference rod NR 4 (Fig. S7); excursions beyond this band mark genuine
reshaping, while signals within it are indistinguishable from a stable
rod.

Figure 3a shows eight consecutive frames of NR 6 alongside the
corresponding asymmetry values, linking each point in the trace directly
to the underlying morphology: material shifts from one tip to the other
and back, which changes the asymmetry magnitude. Across the full
experiment (Fig. 3b), NR 6 shows large, repeated positive and negative
excursions with 62\% of frames exceeding the noise band (Fig. 3b),
demonstrating sustained, highly dynamic pulse-to-pulse reshaping. The
signal alternates sign throughout, showing that the reshaping is at
least partially reversible, although the recurrence is irregular and no
dominant frequency emerges. Rather than progressing towards a single,
more stable morphology, the rod repeatedly revisits configurations it
has already adopted. This successive and repetitive switching forth and
back suggests that, at least over these timescales, successive pulses
are correlated rather than independent: though each pulse is thermally
independent, the configuration left by one pulse appears to bias the
next. It should be noted that NR 6 is one of the most active rods in the
field of view. Across all particles, activity (the fraction of frames
outside the noise band, measured before any mass exchange to be
discussed below) ranges from 1.5\% to 65\%, with a mean of 23\% (Fig.
S8a). The six most active particles (\textgreater40\%) all lie isolated
or at the cluster edge, while the least active (\textless10\%) sit
within it (Fig. S8b). Connectivity thus appears to constrain the
reshaping of individual particles within the cluster.

\begin{figure}[H]
\centering
\includegraphics[width=0.9\textwidth]{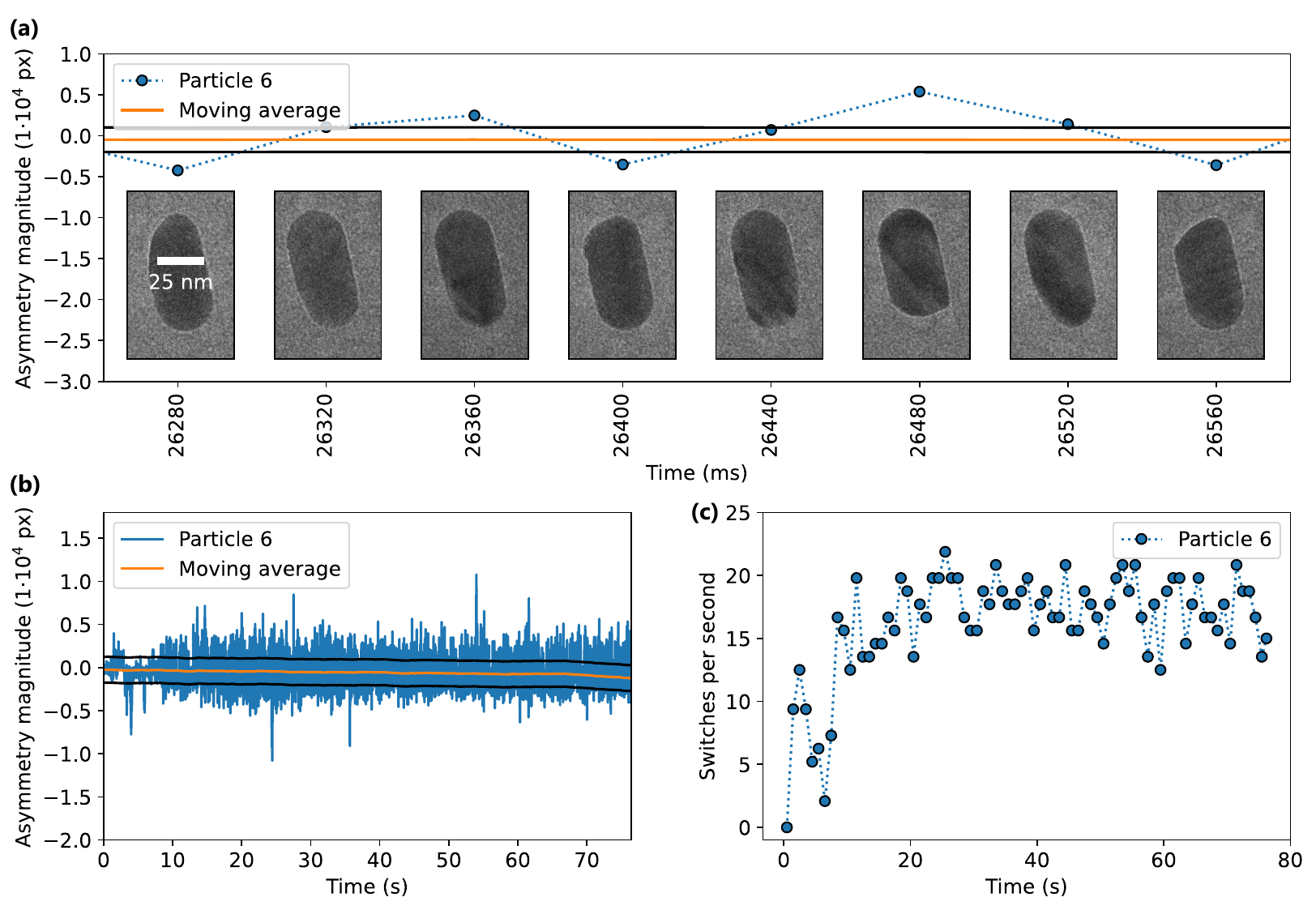}
\caption{Quantitative analysis of pulse-to-pulse
reshaping dynamics in an isolated nanorod. (a) Tip-asymmetry signal
over a short time window showing alternating positive and negative
excursions. Insets show corresponding TEM frames revealing intermediate
morphologies. (b) Tip-asymmetry signal over the full experiment,
demonstrating sustained high-amplitude fluctuations indicating
persistent structural mobility. (c) Switching rate extracted from the
asymmetry signal, showing an initial rise followed by saturation at
\textasciitilde20 switches/s. A switch is counted when the asymmetry
leaves the noise band with the opposite sign to the previous excursion,
that is, when the direction of tip-to-tip transfer reverses by more than
the noise level.}\label{fig3}
\end{figure}

Strikingly, the rate of these sign switches is not constant but evolves
over time: it rises sharply over the first \textasciitilde20 s of
illumination before saturating near 20 switches/s, which translates to
0.8 switches per pulse (Fig. 3c). Because NR 6 is isolated, this
acceleration cannot arise from the growing interparticle connectivity
invoked for the cluster above, and a cumulative thermal origin is
unlikely, given the fast inter-pulse cooling established above. Instead,
it indicates that a comparable ratchet operates within a single
particle: each pulse leaves an irreversible structural increment, in the
local defect population, surface facets or carbon shell, that lowers the
barrier for reshaping by the next, until further pulses add little.
Growing connectivity therefore extends the distance over which gold is
transported but is not required for a progressive increase in reshaping
activity, which occurs in the isolated rod as well. This increase is,
however, not universal: among other active particles, some show a
comparable rise while others do not (Fig. S9). NR 6 shows it most
clearly, and what determines whether it occurs, e.g. differences in
initial connectivity or local environment are plausible factors, remains
unresolved.

\subsection{Coupled reshaping and metastability in a nanorod
pair}\label{coupled-reshaping-and-metastability-in-a-nanorod-pair}

\begin{figure}[H]
\centering
\includegraphics[width=0.9\textwidth]{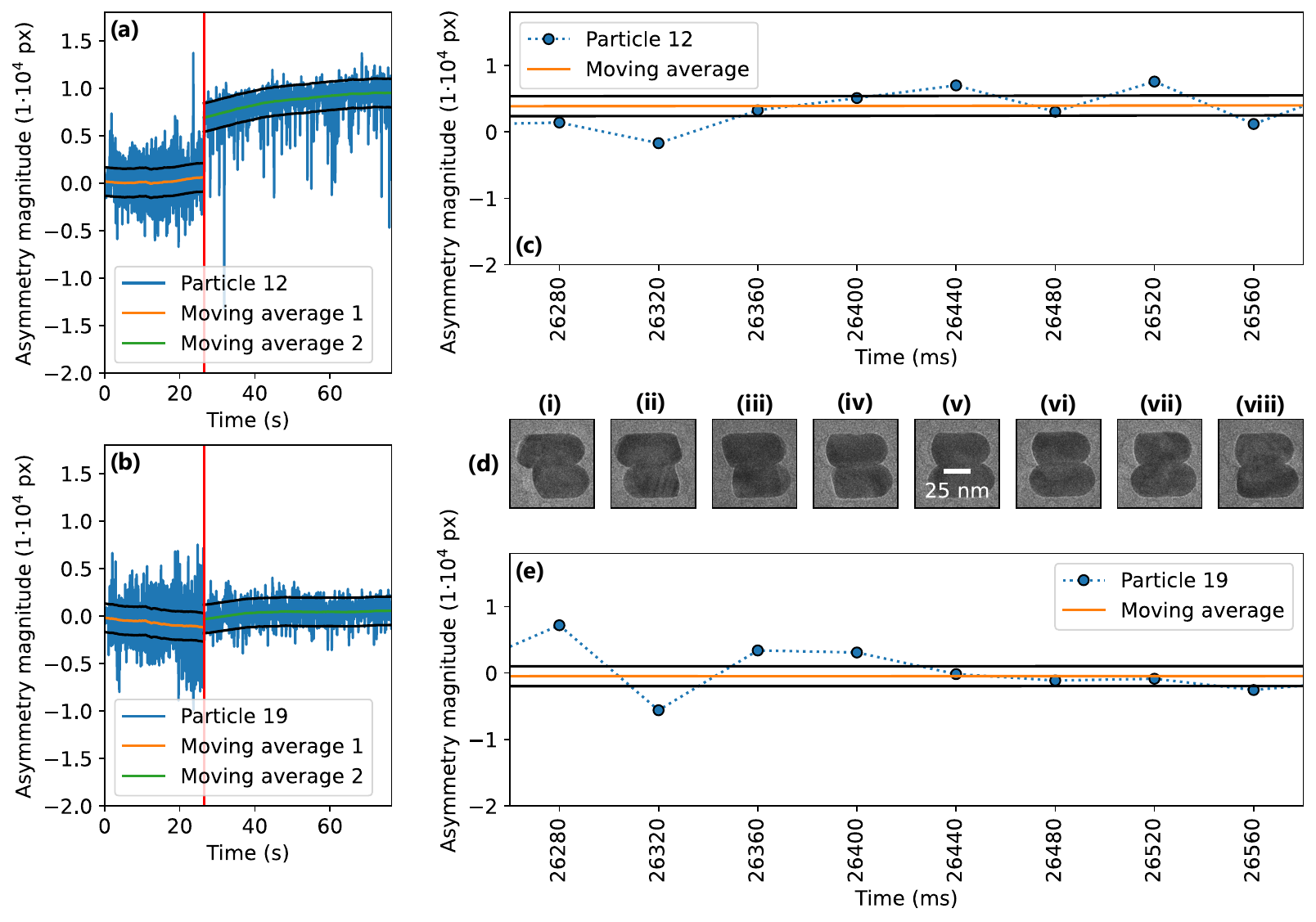}
\caption{Coupled reshaping and dynamical arrest in a
nanorod pair. (a) Full-range evolution of the NR 12 asymmetry signal,
showing a transition to dynamical arrest following a mass-transfer event
(red line). (b) Full-range asymmetry signal for NR 19 displaying the
same immobilization behavior. (c) Short-window tip-asymmetry signal for
NR 12, revealing rapid directional reshaping. (d) TEM zoom-ins of NR 12
and NR 19 corresponding to the times highlighted in panel (c). (e)
Short-window tip-asymmetry signal for NR 19 showing correlated dynamics
with NR 12}\label{fig4}
\end{figure}

NR 6 reshapes in isolation, exchanging no material with its neighbours.
Given the growing connectivity established above, a natural question is
what happens once two rods lie close enough to exchange mass directly.
To probe this, we turn to NR 12 and NR 19, a close pair well separated
from the rest of the cluster. Throughout the experiment up to t = 26.44\,s (red line in Fig. 4a,b), both rods are highly dynamic, each showing
frequent pulse-to-pulse asymmetry switching much like the isolated NR 6.

At t = 26.44\,s this dynamic regime ends abruptly, seen as a sharp
transition in the asymmetry magnitude (red line in Fig. 4a,b). Before
the transition, 58\% of frames for NR 12 and 65\% for NR 19 lie outside
the noise band; afterwards these fall to 17\% and 16\%. Isolated large
excursions persist after the transition, particularly for NR 12, but
these arise from occasional segmentation failures rather than genuine
large reshaping. The same transition appears as a drop in switching rate
for both rods (Fig. S10b,d). This sharp, lasting transition coincides
with a substantial exchange of material between the two rods: NR 12
loses a substantial fraction of its projected area to NR 19 (Fig. S10).
Figure 4d shows the frames spanning the transition (26.28 - 26.56\,s,
bracketing the red line in Fig. 4a,b), with the corresponding difference
maps in Fig. S11b. In the two frames before the exchange, material still
redistributes within the individual rods; the transfer between them then
follows within one to two frames.

This permanent arrest stands in sharp contrast to the isolated rod: NR 6
continues to reshape throughout the experiment, whereas the NR 12 - NR 19
pair drops to near-baseline activity at the moment of mass transfer and
does not recover. The coincidence of a single exchange event with a
lasting transition from sustained mobility to quiescence suggests that
the redistribution of material between the two particles brings the pair
into a configuration that is stable against further excitation, one that
the isolated rod never reaches.

\subsection{Long-range coupling and reversible
disappearance--reappearance
cycles}\label{long-range-coupling-and-reversible-disappearancereappearance-cycles}

Within the dense cluster, reshaping extends well beyond neighbouring
particles. NR 17 and NR 18, separated by several intervening rods at a
centre-to-centre distance of hundreds of nanometres, reshape in tight
synchrony over the same few frames (Fig. 5a). Their projected-area
traces evolve in opposition: as one rod loses footprint, the other gains
it (Fig. 5b). The coupling is most striking in a recurring sequence in
which one rod disappears almost entirely from projection while the other
recovers, passing through an intermediate frame in which both rods show
roughly half their initial contrast at full shape. Each such cycle spans
several frames, markedly slower than the pulse-to-pulse switching of
isolated rods. Molten or otherwise highly mobile gold would appear at
reduced contrast, but such a state exists only during the 400\,$\mu$s pulse,
which occupies \textasciitilde1\% of each frame. Every image is
therefore dominated by the cold, static configuration between pulses, so
the reduced contrast reflects material that has left the site rather
than material that is transiently mobile there. Its magnitude is also
far too large to arise from a change in crystallographic orientation.
The reduced contrast therefore reflects a genuine loss of projected
material at that site. This partial contrast means that only part of the
rod\textquotesingle s material is transferred per pulse, so the exchange
completes over consecutive pulses. Difference maps of the intervening
particles (Fig. S11c) show no detectable accumulation or depletion at any
point in the cycle, indicating that the gold is channelled through to
its destination rather than being deposited and re-released along the
way. The redistribution likely occurs as rapid surface-diffusion during
the hot excitation windows, with the amount passing through any single
intervening rod at any time thin enough to remain undetectable.

\begin{figure}[H]
\centering
\includegraphics[width=0.9\textwidth]{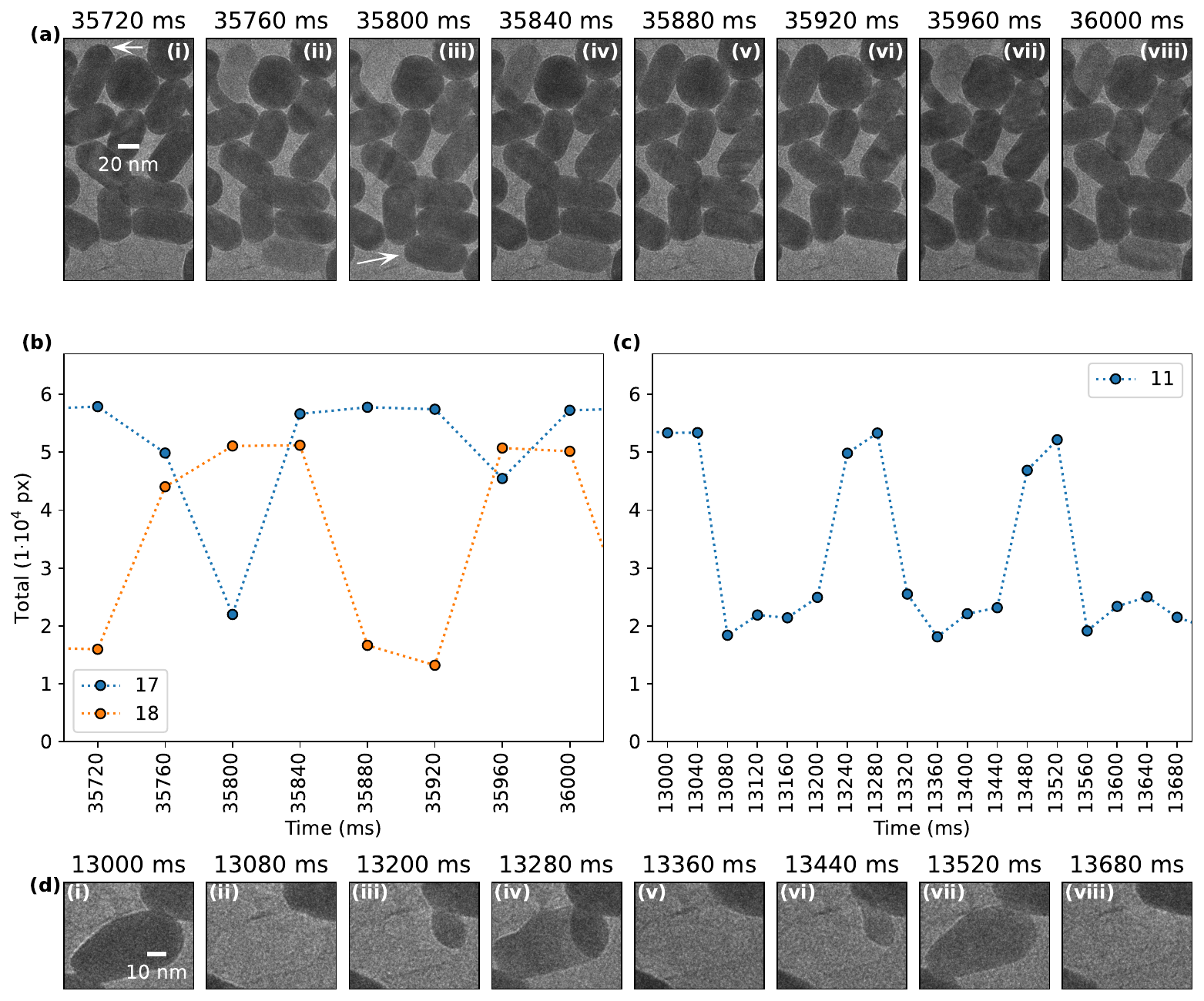}
\caption{Long-range coupled reshaping and reversible
disappearance \textbf{--} reappearance dynamics. (a) TEM zoom-ins
showing two widely separated nanorods (NR 17 and NR 18) undergoing
coordinated reshaping despite multiple intervening rods. (b) Total
projected-area traces for NR 17 and NR 18 over time, revealing
complementary behavior indicative of long-range mass redistribution. (c)
Projected-area trace for NR 11, exhibiting repeated disappearance and
reappearance. (d) Corresponding TEM zoom-ins showing intermediate states
and the robust reformation of the original rod-like morphology.}\label{fig5}
\end{figure}

NR 11 displays the same behaviour in a configuration with no obvious
counterpart. Positioned between two rods and connected to each at its
tips, it repeatedly disappears almost completely from projection and
then reconstitutes itself (Fig. 5c,d), over at least two consecutive
cycles. Reconstructed across frames, the dissolution and recovery
resemble the flow of a liquid, with rounded, partially filled
intermediate shapes; we emphasise that this refers to the apparent
motion, not to the state of any individual frame, which, as established
above, records the cold, solid configuration between pulses. As with NR
17 and NR 18, no neighbouring rod shows an obvious complementary change,
suggesting that the cycling material is exchanged with the two connected
neighbours through the same chain-mediated, multi-pulse redistribution.
Most strikingly, after each near-complete loss of projected material the
rod recovers the same shape, dimensions, and position as before. The
exchange also appears directional: although NR 11 is connected to
neighbouring rods on both sides, material is lost towards and recovered
from the same upper neighbour in both cycles, with a darker neck visible
at that junction (Fig. 5d(iii), (iv) and (vi)). This exact recovery,
along its preferred pathway, shows that the system does not merely
return to some rod-like state but repeatedly reoccupies one specific
metastable configuration, even after passing through configurations in
which almost no material is visible.

The observations across isolated, paired, and clustered nanorods point
consistently to a kinetically controlled, non-equilibrium regime. At the
simulated peak temperature of \textasciitilde1115\,K, below the bulk
melting point of gold but within the range of rapid surface diffusion,
each 400\,$\mu$s pulse opens a high-temperature window orders of magnitude
longer than the picosecond-to-nanosecond timescales of nanoscale
rearrangement. Atoms can therefore explore many configurations during a
pulse and are quenched into a metastable state on cooling. The
persistence of pulse-to-pulse reshaping, rather than a stepwise
relaxation toward equilibrium, indicates that the system is repeatedly
driven through this explore-and-quench cycle. That specific
configurations recur, and that connectivity accumulates, indicates two
further ingredients: a bias that selects particular morphologies on
requench, which appears as structural memory, and the irreversibility of
contact formation between neighbouring particles. A comparable ratchet
appears to operate at two scales: within individual particles, where
accumulated structural changes progressively lower the barrier for
reshaping, and between them, where contact formation extends the
distance over which material is redistributed while constraining the
reshaping of individual particles within the cluster. This internal
ratchet is clearest in the isolated NR 6 and is not seen in every
particle (Fig. S9). Within the cluster, connected particles are
constrained by their neighbours, and material is preferentially
redistributed over long distances rather than by reshaping individual
particles, since breaking an established contact would oppose the same
driving force that formed it.

The distances over which material is redistributed implicate surface,
rather than bulk diffusion. An order-of-magnitude estimate at
\textasciitilde1115\,K gives a bulk self-diffusion length below
\textasciitilde10\,nm per pulse (SI), far too short to connect separate
particles. Surface diffusion is far faster: reported values for
single-adatom hopping on Au(001) and Au(111) give activation energies of
0.67 - 0.84\,eV and diffusion prefactors of 2x10\textsuperscript{-4} -
1x10\textsuperscript{-2}\,cm$^2$s$^{-1}$ \cite{jiang2024crystal}. At our peak temperature,
these parameters give surface diffusion lengths of 100-360\,nm within a
single pulse (SI), comfortably exceeding the \textasciitilde70\,nm
tip-to-tip distance in NR 6 and spanning the \textasciitilde200\,nm NR
17-18 separation. That the NR 17-18 exchange nonetheless needs more
than one pulse therefore reflects how much gold the connecting path can
carry per pulse rather than how far it can travel. These estimates
assume the particles remain solid; any local disordering during the
pulse would increase the transport rate further, so they represent a
lower bound on the accessible distances.

A carbon shell on the particle surfaces, distinct from the amorphous
carbon support and formed from ligand and contamination carbonization
under combined beam and laser exposure, appears central to the
reversibility. In NR 11 we directly observe darker regions at the edges
of such a shell (Fig. 5d), confirming its presence and suggesting that
residual gold remains pinned there between cycles. Because Au-Au
bonding dominates over Au-C \cite{batson2008motion,rodriguez2010trapping,kumar2018wettability}, such residual gold provides a
strong thermodynamic driving force for re-compaction: upon cooling,
dispersed material dewets and reconstitutes into compact, gold-rich
morphologies. The shell thus plausibly both limits sublimation,
consistent with the modest net mass loss in projection across the
ensemble, and biases recovery toward rod-like rather than the
equilibrium Wulff shape. The darker neck at the junction through which
NR 11 empties and refills further suggests that gold wets the carbon
along a preferred path, which would channel the transfer and bias
recovery toward the same configuration. Progressive shell formation is
also a plausible origin for the internal ratchet inferred above,
increasingly templating the recovered shape; whether it develops on the
timescale of the observed incubation (Fig. 3c) remains to be tested. We
cannot resolve the shell\textquotesingle s detailed structure from these
data, so this picture remains an interpretation; one that future
experiments can test through controlled surface modification,
carbon-free supports, varied pulse lengths, and direct temperature
measurement.

\section{Conclusion}\label{conclusion}

Using pulse-synchronized in situ TEM, we resolved the structural
response of gold nanorods to individual optical excitation events.
Because each frame corresponds to one known pulse, the resulting series
preserves the order in which configurations are visited, which no
accumulation of independent before-and-after measurements can provide.
Capturing the structural state after every pulse yields thousands of
frames per experiment; to exploit this, we developed an analysis
framework, including frame registration, segmentation, and
noise-calibrated metrics such as tip asymmetry and switching rate, that
converts the raw image series into quantitative, time-resolved reshaping
trajectories. Together, the acquisition scheme and analysis pipeline
turn laser-driven reshaping into a measurable, per-pulse and
per-particle quantity.

Applying this approach one laser pulse at a time, we uncovered a
surprisingly dynamic regime in which the projected shape of the rods
changes from pulse to pulse, sometimes subtly and sometimes through
large excursions. These changes are at least partly reversible across
scales: an isolated rod shifts material back and forth between its tips,
while rods in the cluster can lose nearly all of their projected
material and later recover the same shape. As discussed above, this
kinetically controlled behaviour reflects rapid surface transport biased
by interfacial constraints rather than relaxation toward equilibrium.
Most striking is the resulting structural memory: that a rod repeatedly
recovers the same shape suggests that specific configurations are not
merely revisited by chance but actively selected.

More broadly, these results establish pulsed \emph{in situ} TEM as a
platform for dissecting non-equilibrium atomic processes one excitation
event at a time. By making mass transport, defect dynamics, and
crystallographic evolution quantitatively trackable under controlled
optical excitation, the approach opens a route to studying and
ultimately steering nanoscale materials whose optical, catalytic, or
mechanical functions emerge from transient or repeatedly accessed
structural states. Controlling such configurations through the timing
and intensity of optical stimuli offers a possible path toward
stimulus-responsive nanomaterials and time-programmable catalytic
systems, in which transient geometries, defect populations, or
electronic states are selectively populated during operation.

\section{Methods}

\subsection{Experimental}

Gold nanorods (Au NRs; 40 nm × 68\,nm, NanoSeedz Ltd.) were deposited
onto amorphous carbon TEM grids and excited in situ inside a JEOL
GrandARM TEM (300 kV) equipped with an IDES Luminary Micro
laser-incoupling system (Fig. 1a). A 577\,nm modulated CW laser was
focused to a 40\,$\mu$m spot to excite the nanorods near their simulated
longitudinal LSPR at 580 nm (Fig. S1). The modulated CW laser delivered
400 µs optical pulses synchronized to the 40\,ms camera frame, so that
each frame captures the structural state of exactly one pulse. The
relative phase between the laser pulse train and the camera exposure was
unfortunately not retrievable in this experiment. A pulse falling within
the exposure would average the configurations before and after it, so
that any particle reshaping during that pulse would appear at partial
contrast in both configurations. NR 6 reverses its tip asymmetry on the
majority of pulses, with tip area changes that are clearly resolved in
individual frames, yet these frames show a single sharply defined
morphology rather than two superimposed tip configurations. We therefore
infer that the pulse falls close to a frame boundary, so that each frame
is dominated by the structural state following a single pulse.

\subsection{Electrodynamic simulations}

BEM simulations were performed using a GPU-accelarated version of the
MNPBEM17 Matlab toolbox. A retarded solver was used (so not
quasi-static) that interprets the surfaces as curved. The trirod
function in the toolbox was used to create a rod with a diameter and
length of 40x68 nm, respectively. We used our recommended meshing sizes
of $dis_{phi} \approx3$
nm, $dis_{theta}\approx1$\,nm, and
$dis_{z}\approx3$\,nm to create a triangular mesh. A
spectral range of 400 to 900\,nm was used with steps of 2\,nm and we
excited the system with a planewave propagating to -z with polarization
along x (in this case the long axis of our rod). For the dielectric
function of the particle we used the data from McPeak et al. For the
dielectric function of the environment we compared n=1 (vacuum) to an
effective medium approximation using Bruggeman\textquotesingle s model:

\begin{equation}
     \epsilon_{eff}= \frac{H_b+\sqrt{H_b^2+8\epsilon_1\epsilon_2}}{4}
\end{equation}

\begin{center}
with
\end{center}
\[H_b=(3c_1-1)e_1+(3c_2-1)e_2\]

where we used n=2.4 for the carbon film and a ratio of 1:9
vacuum:carbon, such that $e_{eff}=1.32$. The results from these
simulations are depicted in Figure S1.

\subsection{Absorbed power in the carbon substrate}

To quantify the optical loading of the support membrane, we simulated
absorption of the 577\,nm laser light within the amorphous carbon TEM
film using the electromagnetic-wave frequency-domain module in COMSOL
Multiphysics v5.5. The domain consisted of a 20\,nm-thick carbon membrane
suspended in vacuum and illuminated by a linearly polarized plane wave
incident along the surface normal. Material parameters were obtained
from Larruquert et al. \cite{larruquert2013high}. A perfectly matched layer of
thickness 2\,$\mu$m was applied at the boundaries to prevent reflection
artefacts.

Because the illuminated region spans tens of micrometres, but absorption
within the thin film is homogeneous at this wavelength, we simulated an
8\,$\mu$m x 8\,$\mu$m subsection to reduce computation time. The volumetric
absorbed power was calculated by averaging the local power dissipation
within the film volume, yielding an absorbed power density of 3.17 x
10\textsuperscript{13} W/m\textsuperscript{3}, which we used as a
spatially Gaussian heat source when computing the temperature rise of
the TEM grid (main-text Fig. 1e). Plasmonic heating turned out to be
inferior. Nonetheless we simulated the expected absorption in
Supplementary Fig. S1.

\subsection{Heat-diffusion simulations}

We modelled heat diffusion within a full 100\,$\mu$m x 100\,$\mu$m TEM-window
region containing the 43 particles extracted from Supplementary Fig. S4.
The geometry of each sphere and rod, its footprint, length, and aspect
ratio, was taken directly from the mask fits applied during automated
segmentation. The simulation domain comprised the carbon support and the
nanorods, both surrounded by vacuum.

Heat diffusion was computed using COMSOL's Heat Transfer in Solids
module. Material properties are summarized in Tab. S. At the
temperatures reached in the steady-state simulation, amorphous carbon
partially crystallizes toward a turbostratic graphite-like structure.
Because strongly twisted graphite layers exhibit dramatically reduced
in‑plane conductivity compared to single-crystal graphite, we used a
conservative effective in-plane thermal conductivity of 10\,W/m$\cdot$K,
consistent with prior studies on thin, disordered graphite layers.

\textbf{Tab.} \textbf{1} Parameters used for the heat simulations.

\begin{longtable}[]{@{}
  >{\raggedright\arraybackslash}p{(\linewidth - 2\tabcolsep) * \real{0.5}}
  >{\raggedright\arraybackslash}p{(\linewidth - 2\tabcolsep) * \real{0.5}}@{}}
\toprule\noalign{}
\begin{minipage}[b]{\linewidth}\raggedright
Thermal conductivity of gold
\end{minipage} & \begin{minipage}[b]{\linewidth}\raggedright
315 W/m$\cdot$K
\end{minipage} \\
\midrule\noalign{}
\endhead
\bottomrule\noalign{}
\endlastfoot
Heat capacity of gold & 129\,J/kg$\cdot$K \\
Density of gold & 19320\,kg/m\textsuperscript{3} \\
Thermal conductivity of carbon & 10\,W/m$\cdot$K \\
Heat capacity of carbon & 710\,J/kg$\cdot$K \\
Density of carbon & 2330\,kg/m\textsuperscript{3} \\
Laser beam radius & 20\,$\mu$m \\
Laser power & 50\,mW \\
\end{longtable}

Nanorods were assigned spatially uniform heat sources corresponding to
the product of their optical absorption cross-sections and the laser
fluence. The carbon support was heated using the Gaussian profile
derived from the absorbed-power calculation described above, with a 40\,$\mu$m radius corresponding to the experimental beam size. Fixed-temperature
boundary conditions were applied at the outer edges of the membrane,
representing the copper grid that acts as a heat sink.

The domain was meshed with tetrahedral elements of size 1-10\,nm and
solved using a fully coupled GMRES solver to obtain the steady-state
temperature distribution shown in main-text Fig. 1e.
\newpage
\bibliography{sn-bibliography}

@article{dubau2016defects,
  title={Defects do catalysis: CO monolayer oxidation and oxygen reduction reaction on hollow PtNi/C nanoparticles},
  author={Dubau, Laetitia and Nelayah, Jaysen and Moldovan, Simona and Ersen, Ovidiu and Bordet, Pierre and Drnec, Jakub and Asset, Tristan and Chattot, Rapha{\"e}l and Maillard, Fr{\'e}d{\'e}ric},
  journal={ACS Catalysis},
  volume={6},
  number={7},
  pages={4673--4684},
  year={2016},
  publisher={ACS Publications}
}

@article{jia2019role,
  title={The role of defect sites in nanomaterials for electrocatalytic energy conversion},
  author={Jia, Yi and Jiang, Kun and Wang, Haotian and Yao, Xiangdong},
  journal={Chem},
  volume={5},
  number={6},
  pages={1371--1397},
  year={2019},
  publisher={Elsevier}
}

@article{lak2021embracing,
  title={Embracing defects and disorder in magnetic nanoparticles},
  author={Lak, Aidin and Disch, Sabrina and Bender, Philipp},
  journal={Advanced Science},
  volume={8},
  number={7},
  pages={2002682},
  year={2021},
  publisher={Wiley Online Library}
}

@article{oliveira20203d,
  title={3D Characterization and plasmon mapping of gold nanorods welded by femtosecond laser irradiation},
  author={Oliveira, MT and Albrecht, W and Gonz{\'a}lez-Rubio, G and Altantzis, T and Hoyos, LIP and B{\'e}ch{\'e}, A and Aert, SV and Guerrero-Mart{\'\i}nez, A and Liz-Marz{\'a}n, LM and Bals, S},
  journal={ACS Nano},
  volume={14},
  number={10},
  pages={12558--12570},
  year={2020}
}

@article{albrecht20213d,
  title={3D Atomic-Scale Dynamics of Laser-Light-Induced Restructuring of Nanoparticles Unraveled by Electron Tomography},
  author={Albrecht, Wiebke and Arslan Irmak, Ece and Altantzis, Thomas and Pedrazo-Tardajos, Adri{\'a}n and Skorikov, Alexander and Deng, Tian-Song and van der Hoeven, Jessi ES and van Blaaderen, Alfons and Van Aert, Sandra and Bals, Sara},
  journal={Advanced Materials},
  volume={33},
  number={33},
  pages={2100972},
  year={2021},
  publisher={Wiley Online Library}
}

@article{gonzalez2021controlled,
  title={Controlled Alloying of Au@ Ag Core--Shell Nanorods Induced by Femtosecond Laser Irradiation},
  author={Gonz{\'a}lez-Rubio, Guillermo and D{\'\i}az-N{\'u}{\~n}ez, Pablo and Albrecht, Wiebke and Manzaneda-Gonz{\'a}lez, Vanesa and Ba{\~n}ares, Luis and Rivera, Antonio and Liz-Marz{\'a}n, Luis M and Pe{\~n}a-Rodr{\'\i}guez, Ovidio and Bals, Sara and Guerrero-Mart{\'\i}nez, Andr{\'e}s},
  journal={Advanced Optical Materials},
  volume={9},
  number={10},
  pages={2002134},
  year={2021},
  publisher={Wiley Online Library}
}

@article{gonzalez2017femtosecond,
  title={Femtosecond laser reshaping yields gold nanorods with ultranarrow surface plasmon resonances},
  author={Gonz{\'a}lez-Rubio, Guillermo and D{\'\i}az-N{\'u}{\~n}ez, Pablo and Rivera, Antonio and Prada, Alejandro and Tardajos, Gloria and Gonz{\'a}lez-Izquierdo, Jes{\'u}s and Ba{\~n}ares, Luis and Llombart, Pablo and Macdowell, Luis G and Alcolea Palafox, Mauricio and others},
  journal={Science},
  volume={358},
  number={6363},
  pages={640--644},
  year={2017},
  publisher={American Association for the Advancement of Science}
}

@article{manzaneda2023multi,
  title={From multi-to single-hollow trimetallic nanocrystals by ultrafast heating},
  author={Manzaneda-Gonz{\'a}lez, Vanesa and Jenkinson, Kellie and Pe{\~n}a-Rodr{\'\i}guez, Ovidio and Borrell-Grueiro, Olivia and Trivi{\~n}o-S{\'a}nchez, Sergio and Ba{\~n}ares, Luis and Junquera, Elena and Espinosa, Ana and Gonz{\'a}lez-Rubio, Guillermo and Bals, Sara and others},
  journal={Chemistry of Materials},
  volume={35},
  number={22},
  pages={9603--9612},
  year={2023},
  publisher={ACS Publications}
}

@article{milagres20203d,
  title={3D characterization and plasmon mapping of gold nanorods welded by femtosecond laser irradiation},
  author={Milagres de Oliveira, Thaís and Albrecht, Wiebke and Gonz{\'a}lez-Rubio, Guillermo and Altantzis, Thomas and Lobato Hoyos, Ivan Pedro and B{\'e}ch{\'e}, Armand and Van Aert, Sandra and Guerrero-Mart{\'\i}nez, Andr{\'e}s and Liz-Marzan, Luis M and Bals, Sara},
  journal={ACS nano},
  volume={14},
  number={10},
  pages={12558--12570},
  year={2020},
  publisher={ACS Publications}
}

@article{gonzalez2020formation,
  title={Formation of hollow gold nanocrystals by nanosecond laser irradiation},
  author={Gonz{\'a}lez-Rubio, Guillermo and Milagres de Oliveira, Thais and Albrecht, Wiebke and D{\'\i}az-N{\'u}{\~n}ez, Pablo and Castro-Palacio, Juan Carlos and Prada, Alejandro and Gonz{\'a}lez, Rafael I and Scarabelli, Leonardo and Ba{\~n}ares, Luis and Rivera, Antonio and others},
  journal={The journal of physical chemistry letters},
  volume={11},
  number={3},
  pages={670--677},
  year={2020},
  publisher={ACS Publications}
}

@article{gonzalez2016reshaping,
  title={Reshaping, fragmentation, and assembly of gold nanoparticles assisted by pulse lasers},
  author={Gonz{\'a}lez-Rubio, Guillermo and Guerrero-Mart{\'\i}nez, Andr{\'e}s and Liz-Marz{\'a}n, Luis M},
  journal={Accounts of chemical research},
  volume={49},
  number={4},
  pages={678--686},
  year={2016},
  publisher={ACS Publications}
}

@article{gonzalez2015femtosecond,
  title={Femtosecond laser-controlled tip-to-tip assembly and welding of gold nanorods},
  author={Gonz{\'a}lez-Rubio, Guillermo and Gonz{\'a}lez-Izquierdo, Jes{\'u}s and Ba{\~n}ares, Luis and Tardajos, Gloria and Rivera, Antonio and Altantzis, Thomas and Bals, Sara and Pe{\~n}a-Rodr{\'\i}guez, Ovidio and Guerrero-Mart{\'\i}nez, Andr{\'e}s and Liz-Marz{\'a}n, Luis M},
  journal={Nano letters},
  volume={15},
  number={12},
  pages={8282--8288},
  year={2015},
  publisher={ACS Publications}
}

@article{zak2022light,
  title={Light-Induced In Situ Transmission Electron Microscopy -- Development, Challenges, and Perspectives},
  author={Zak, Andrzej M},
  journal={Nano Letters},
  volume={22},
  number={23},
  pages={9219--9226},
  year={2022},
  publisher={ACS Publications}
}

@article{castro2020hollow,
  title={Hollow gold nanoparticles produced by femtosecond laser irradiation},
  author={Castro-Palacio, Juan Carlos and Ladutenko, Konstantin and Prada, Alejandro and Gonz{\'a}lez-Rubio, Guillermo and D{\'\i}az-N{\'u}{\~n}ez, Pablo and Guerrero-Mart{\'\i}nez, Andr{\'e}s and Fern{\'a}ndez de C{\'o}rdoba, Pedro and Kohanoff, Jorge and Perlado, Jos{\'e} Manuel and Pe{\~n}a-Rodr{\'\i}guez, Ovidio and others},
  journal={The journal of physical chemistry letters},
  volume={11},
  number={13},
  pages={5108--5114},
  year={2020},
  publisher={ACS Publications}
}

@article{diaz2018using,
  title={Using femtosecond laser irradiation to grow the belly of gold nanorods},
  author={D{\'\i}az-N{\'u}{\~n}ez, Pablo and Gonz{\'a}lez-Rubio, Guillermo and Prada, Alejandro and Gonz{\'a}lez Izquierdo, Jesús and Rivera, Antonio and Ba{\~n}ares, Luis and Guerrero-Mart{\'\i}nez, Andr{\'e}s and Pe{\~n}a-Rodr{\'\i}guez, Ovidio},
  journal={The Journal of Physical Chemistry C},
  volume={122},
  number={34},
  pages={19816--19822},
  year={2018},
  publisher={ACS Publications}
}

@article{gonzalez2022engineering,
  title={Engineering of plasmonic gold nanocrystals through pulsed laser irradiation},
  author={Gonz{\'a}lez-Rubio, Guillermo and Albrecht, Wiebke},
  journal={Applied Physics Letters},
  volume={121},
  number={20},
  year={2022},
  publisher={AIP Publishing}
}

@article{link2000does,
  title={How does a gold nanorod melt?},
  author={Link, Stephan and Wang, Zhong L and El-Sayed, Mostafa A},
  journal={The Journal of Physical Chemistry B},
  volume={104},
  number={33},
  pages={7867--7870},
  year={2000},
  publisher={ACS Publications}
}

@article{opletal2011stability,
  title={Stability and transformations of heated gold nanorods},
  author={Opletal, George and Grochola, Gregory and Chui, Yu Hang and Snook, IK and Russo, SP},
  journal={The Journal of Physical Chemistry C},
  volume={115},
  number={11},
  pages={4375--4380},
  year={2011},
  publisher={ACS Publications}
}

@article{wang2005surface,
  title={Surface-driven bulk reorganization of gold nanorods},
  author={Wang, Yanting and Teitel, S and Dellago, Christoph},
  journal={Nano letters},
  volume={5},
  number={11},
  pages={2174--2178},
  year={2005},
  publisher={ACS Publications}
}

@article{yan2020light,
  title={Light tailoring of internal atomic structure of gold nanorods},
  author={Yan, Jianfeng and Zhu, Dezhi and Xie, Jiawang and Shao, Yang and Xiao, Wei},
  journal={Small},
  volume={16},
  number={22},
  pages={2001101},
  year={2020},
  publisher={Wiley Online Library}
}

@article{sumimoto2014situ,
  title={In situ observation of structural transformation of gold nanorods under pulsed laser irradiation in an HVEM},
  author={Sumimoto, Nao and Nakao, Koichiro and Yamamoto, Tomokazu and Yasuda, Kazuhiro and Matsumura, Syo and Niidome, Yasuro},
  journal={Journal of Electron Microscopy},
  volume={63},
  number={4},
  pages={261--268},
  year={2014},
  publisher={The Japanese Society of Microscopy}
}

@article{albrecht2018impact,
  title={Impact of the electron beam on the thermal stability of gold nanorods studied by environmental transmission electron microscopy},
  author={Albrecht, Wiebke and van de Glind, Arjen and Yoshida, Hideto and Isozaki, Yusuke and Imhof, Arnout and van Blaaderen, Alfons and de Jongh, Petra E and de Jong, Krijn P and Ze{\v{c}}evi{\'c}, Jovana and Takeda, Seiji},
  journal={Ultramicroscopy},
  volume={193},
  pages={97--103},
  year={2018},
  publisher={Elsevier}
}

@article{staechelin2021size,
  title={Size-dependent electron--phonon coupling in monocrystalline gold nanoparticles},
  author={Staechelin, Yannic U and Hoeing, Dominik and Schulz, Florian and Lange, Holger},
  journal={ACS Photonics},
  volume={8},
  number={3},
  pages={752--757},
  year={2021},
  publisher={ACS Publications}
}

@article{chiang2023electron,
  title={Electron--phonon relaxation dynamics of hot electrons in gold nanoparticles are independent of excitation pathway},
  author={Chiang, Wei-Yi and Bruncz, Autumn and Ostovar, Behnaz and Searles, Emily K and Brasel, Sadie and Hartland, Gregory and Link, Stephan},
  journal={The Journal of Physical Chemistry C},
  volume={127},
  number={43},
  pages={21176--21185},
  year={2023},
  publisher={ACS Publications}
}

@article{larruquert2013high,
  title={High reflectance ta-C coatings in the extreme ultraviolet},
  author={Larruquert, Juan Ignacio and Rodr{\'\i}guez-de Marcos, LV and M{\'e}ndez, Jos{\'e} Antonio and Martin, PJ and Bendavid, A},
  journal={Optics express},
  volume={21},
  number={23},
  pages={27537--27549},
  year={2013},
  publisher={Optical Society of America}
}

@article{cho2020atomic,
  title={Atomic scale mechanisms underlying thermal reshaping of anisotropic gold nanocrystals revealed by in situ electron microscopy},
  author={Cho, Hoduk and Shin, Jae Won and Ryoo, Ryong},
  journal={The Journal of Physical Chemistry C},
  volume={124},
  number={23},
  pages={12855--12863},
  year={2020},
  publisher={ACS Publications}
}

@article{imura2025ultrafine,
  title={Ultrafine Modulation of Plasmon Resonance in Single Gold Nanorods via Photothermal Effects},
  author={Imura, Kohei and Hayashi, Naoya and Kawashima, Kenya},
  journal={The Journal of Physical Chemistry Letters},
  volume={16},
  pages={11037--11042},
  year={2025},
  publisher={ACS Publications}
}

@article{petrova2006temperature,
  title={On the temperature stability of gold nanorods: comparison between thermal and ultrafast laser-induced heating},
  author={Petrova, Hristina and Juste, Jorge Perez and Pastoriza-Santos, Isabel and Hartland, Gregory V and Liz-Marz{\'a}n, Luis M and Mulvaney, Paul},
  journal={Physical Chemistry Chemical Physics},
  volume={8},
  number={7},
  pages={814--821},
  year={2006},
  publisher={Royal Society of Chemistry}
}

@article{batson2008motion,
  title={Motion of gold atoms on carbon in the aberration-corrected STEM},
  author={Batson, Philip E},
  journal={Microscopy and Microanalysis},
  volume={14},
  number={1},
  pages={89--97},
  year={2008},
  publisher={Oxford University Press}
}

@article{rodriguez2010trapping,
  title={Trapping of metal atoms in vacancies of carbon nanotubes and graphene},
  author={Rodr{\'\i}guez-Manzo, Julio A and Cretu, Ovidiu and Banhart, Florian},
  journal={ACS nano},
  volume={4},
  number={6},
  pages={3422--3428},
  year={2010},
  publisher={ACS Publications}
}

@article{kumar2018wettability,
  title={Wettability and structural evolution of gold over a single-walled carbon nanotube: an atomistic investigation},
  author={Kumar, Sunil},
  journal={The Journal of Physical Chemistry C},
  volume={122},
  number={28},
  pages={16346--16355},
  year={2018},
  publisher={ACS Publications}
}

@article{monai2026grand,
  title={Grand challenges and opportunities in stimulated dynamic and resonant catalysis},
  author={Monai, Matteo and Albrecht, Wiebke and Alkemper, Achim and Artrith, Nongnuch and Baldi, Andrea and Beck, Arik and Berry, Ryan T and Bianco, Ettore and Brzesowsky, Floor A and Dong, Qi and others},
  journal={ACS catalysis},
  volume={16},
  number={5},
  pages={4077--4112},
  year={2026},
  publisher={ACS Publications}
}

@article{wittreich2022catalytic,
  title={Catalytic resonance of ammonia synthesis by simulated dynamic ruthenium crystal strain},
  author={Wittreich, Gerhard R and Liu, Shizhong and Dauenhauer, Paul J and Vlachos, Dionisios G},
  journal={Science advances},
  volume={8},
  number={4},
  pages={eabl6576},
  year={2022},
  publisher={American Association for the Advancement of Science}
}

@article{jiang2024crystal,
  title={Crystal plane orientation-dependent surface atom diffusion in sub--10-nm Au nanocrystals},
  author={Jiang, Junnan and Chu, Shufen and Zhang, Yin and Sun, Guangbin and Jin, Junhui and Zeng, Xiaoqin and Chen, Mingwei and Liu, Pan},
  journal={Science Advances},
  volume={10},
  number={21},
  pages={eadn5946},
  year={2024},
  publisher={American Association for the Advancement of Science}
}



\section*{Supplementary material}

# Supplementary information

## Pulse-by-pulse imaging of reversible, collective reshaping in laser-excited gold nanorods

*Mess Dieperink[1], Loriane Monin[1], Ichiro Ohnishi[2],Takeo Sasaki[2], Yu Jimbo[2], Erik Garnett[1], Wiebke Albrecht[1*]*

[1]Center for Nanophotonics, NWO-Institute AMOLF, Amsterdam, The Netherlands

[2]EM Business Unit, JEOL, Tokyo, Japan

*Email Address: w.albrecht@amolf.nl*

*Optical simulations*

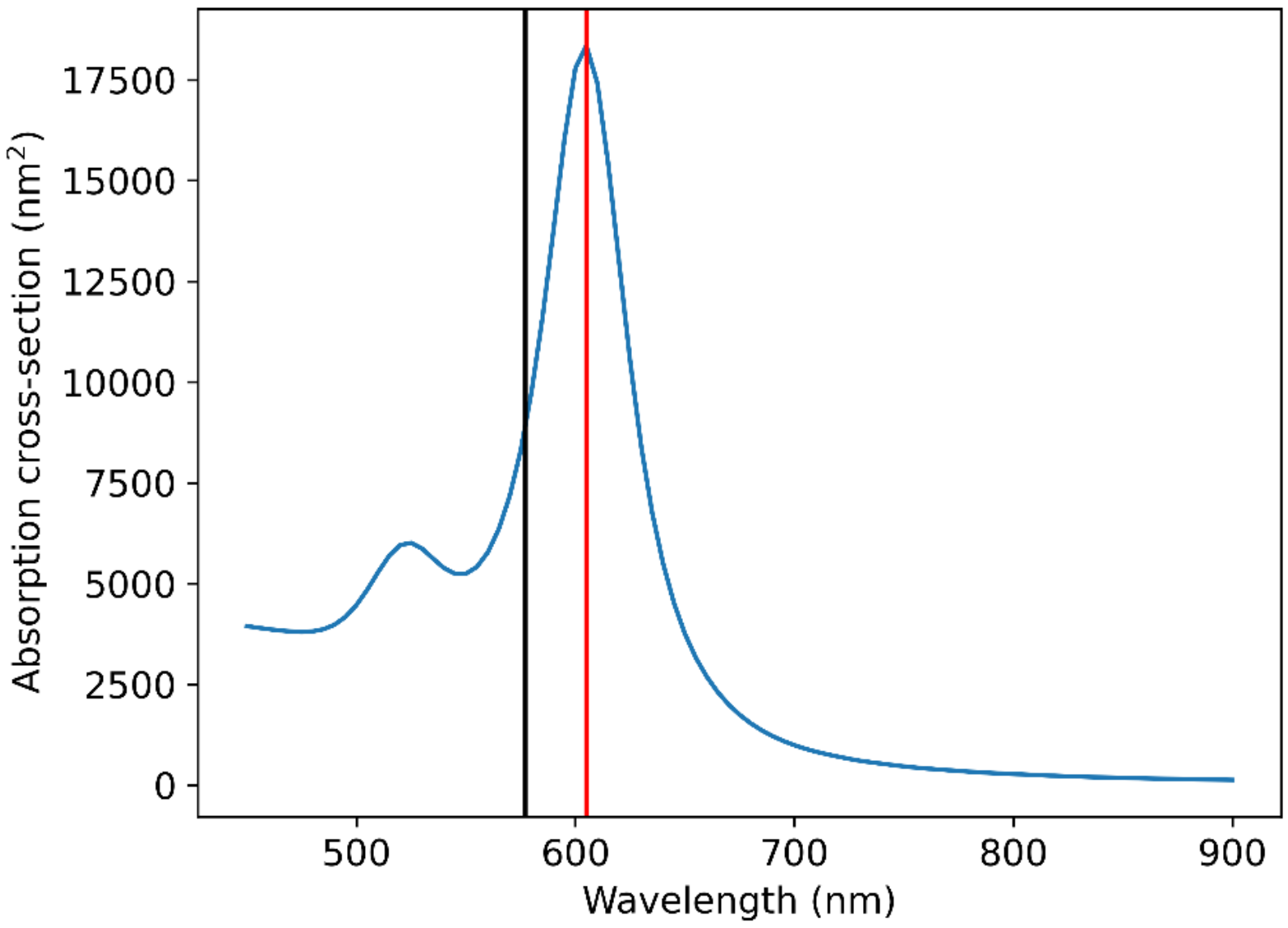


**Fig. S1 Optical absorption characteristics of a gold nanorod.** Absorption cross-section of a 40 x 68 nm AuNR with an effective medium of f = 0.9 for air and carbon. The red line indicates the peak maximum and the black line the laser excitation at 577 nm.

*Mask binarization and manual segmentation*

Manual segmentation was performed on the initial and final frames of the experiment to validate automated image processing. Supplementary Fig. S2a,b show the binarized versions of main-text Fig. 1c,d. Thresholding was performed using Otsu's method, followed by island-size filtering to remove background noise. The segmented masks used in the main text (Supplementary Fig. S2c,d) were generated in ImageJ by manually removing connecting pixels between nanorods using a 2-pixel brush, matching the 2-connectedness used in island identification.

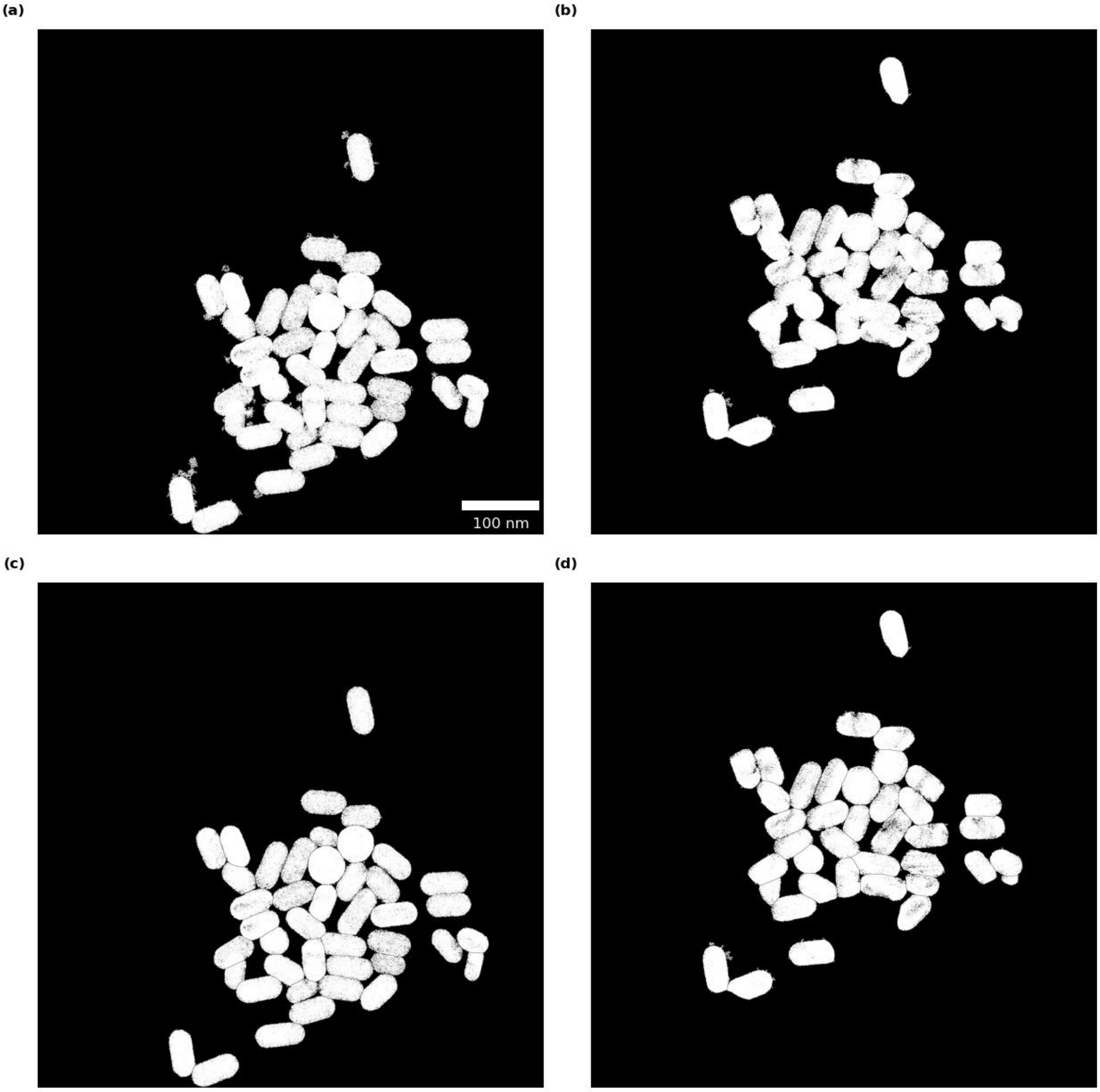


**Fig. S2 Segmentation workflow.** (a,b) Binarized images based on automated Otsu segmentation corresponding to main-text Fig. 1c,d. (c,d) Manually segmented masks obtained using ImageJ.

*Difference map and relative change analysis*

To quantify how much each particle changed between the initial and final state, we compared the overview images recorded before and after excitation (Fig. 1c,d). A true volume change quantification would require tomography of the cluster in both states, which was not feasible here; the analysis below is therefore based on projections and yields an estimate rather than an absolute volume balance. Subtracting the final overview image (Fig. 1d) from the initial one (Fig. 1c) gives the pixel-by-pixel difference in contrast shown in Supplementary Fig. S3a, which identifies where projected material was gained and where it was lost.

The color scale used throughout the Supplementary Information is introduced here:

- **red** = gain in projected area relative to initial configuration,
- **blue**= loss in projected area,
- **white** = unchanged.

For a quantitative measure, each particle was segmented in both images and its projected area obtained by summing the enclosed pixels. Supplementary Fig. S3b shows the relative change in projected area per particle. Most particles gained projected area (red dots in Fig. S3b) with relative positive changes up to 20% in projected area. A few particles lost projected area, the most dominantly the four particles that completely disappeared. Summing all negative and positive changes in projected area results in a net loss of -1.4 × $10^5$ px, which is about the projected area of approximately two nanorods.

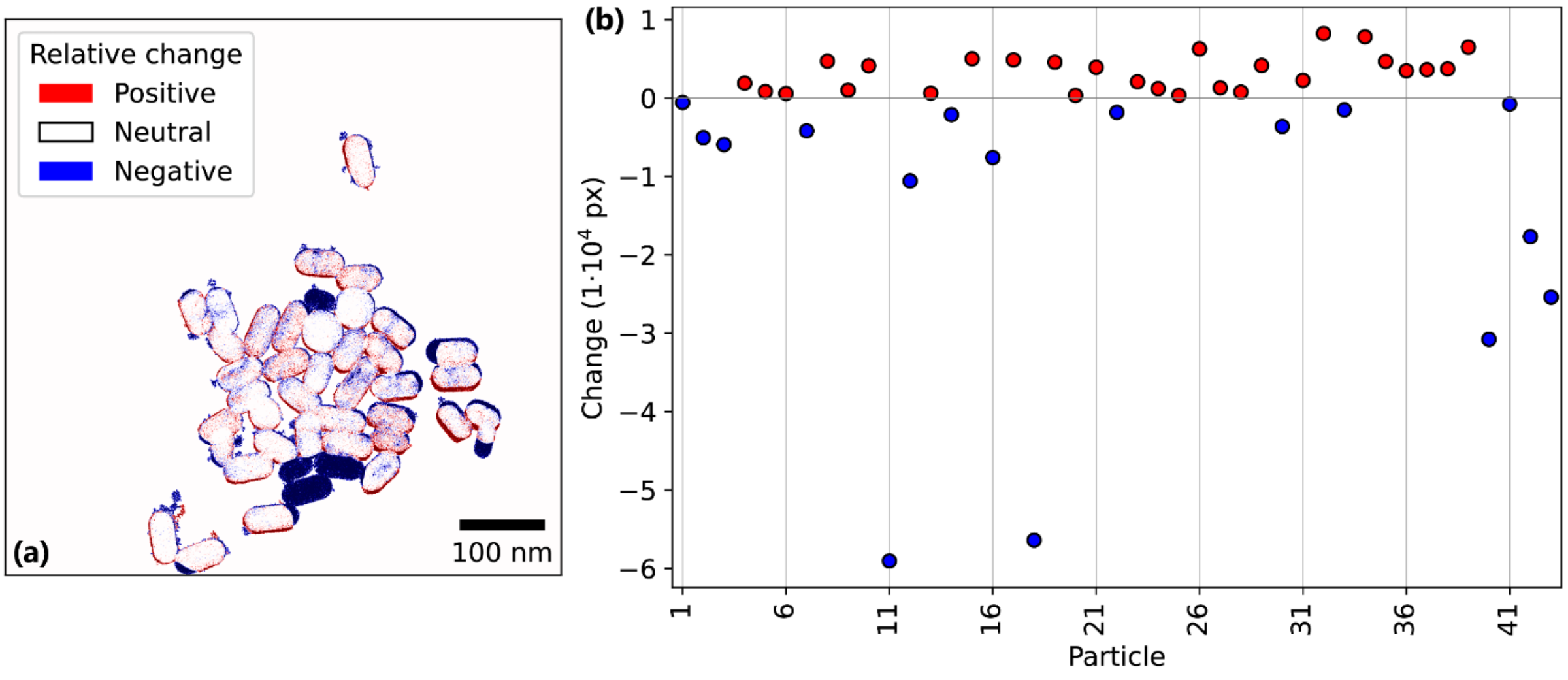


**Fig. S3 Difference-map and absolute-change analysis.** (a) Difference map between the initial and final cluster frames (main-text Fig. 1c,d), using the color scale introduced here. (b) Absolute projected-area change for each particle.

*Cooling dynamics between pulses*

To verify that heating does not accumulate between pulses, we used the steady-state temperature distribution as the initial condition for a time-dependent simulation. After turning off all heat sources, we let the system relax back to equilibrium. The temperature at every nanorod returned to within a few degrees of room temperature within approximately 200 μs (Supplementary Fig. S4). Because this cooling time is much shorter than the 39.6 ms interval between optical pulses, each pulse acts as an independent heating event, consistent with the interpretation presented in the main text.

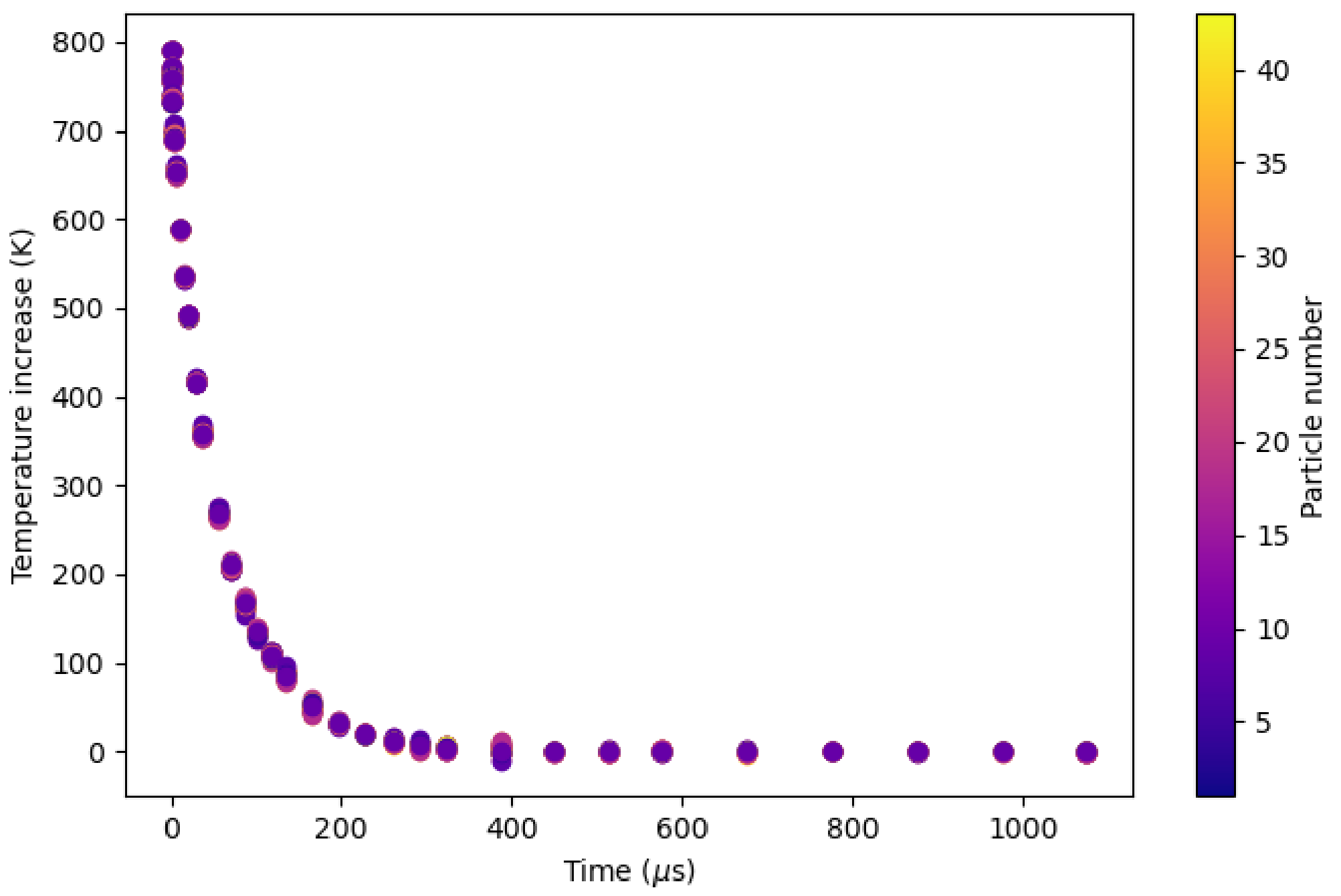


**Fig. S4 Cooling dynamics following a single pulse.** Nanoparticles cool to near room temperature within ~200 μs, ensuring no heat accumulation between pulses. The temperature decay is plotted for every particle; the overlapping curves indicate that there is no spatial gradient in temperature for the complete field of view.

*Enhanced-contrast twinning frames*

Supplementary Fig. S5 provides zoomed-in improved-contrast versions of the twinning-defect frames shown in main-text Fig. 2. Panels (a–c) correspond to the insets of main-text Fig. 2a-c. Panels (d) and (e) show two consecutive enhanced-contrast frames illustrating twin-boundary reorientation.

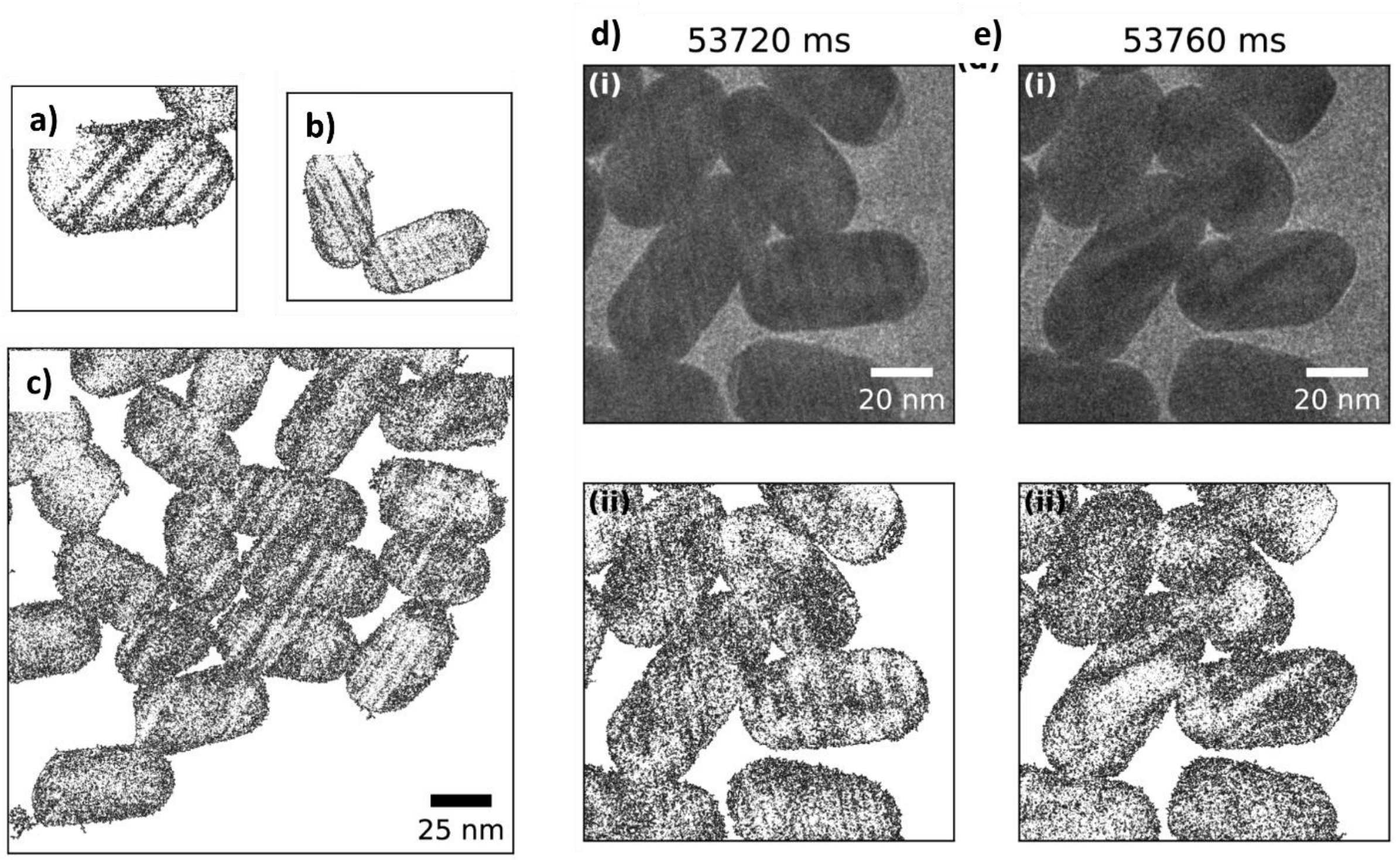


**Fig. S5 Enhanced-contrast twinning-defect frames.** (a–c) High-contrast versions of twinning insets from main-text Fig. 2a-c. (d,e) Consecutive frames showing twin-boundary reorientation.

*Segmentation and masking pipeline*

To extract quantitative dynamics, automated segmentation was performed for all 2398 frames (Fig. S2a,b). Supplementary Fig. S6 summarizes the subsequent process of particle tracking and fitting. All rods were labelled in the manually segmented initial frame and fitted with geometric masks (Fig. S6a). Holes due to diffraction contrast were filled (Fig. S6b). A mask was assigned to each rod with fixed dimensions; translation and rotation were updated each frame using the drift and orientation of NR 6 as reference (Fig. S6c). This was necessary to correct for sample drift during the laser excitation experiment. A shortcoming of this method is that no size increases could be captured because of the fixed size of the mask. However, over the course of the experiments we observed slight shrinking of the frame (Fig. S6d), likely due to laser-induced heating and restructuring of the carbon support. This shrinking partially compensated these mask-size constraints.

For each frame i, the tip-asymmetry $A_i$ is the difference in projected area between the two halves of the rod mask, in pixels. A frame is classified as active when $|A_i|$ exceeds the noise band calibrated on the static reference rod NR 4. A switch is counted when an active frame has the opposite sign to the preceding active frame, so that intervening frames within the noise band do not interrupt the comparison. The switching rate is the number of switches within a sliding window of N frames, divided by the window duration.

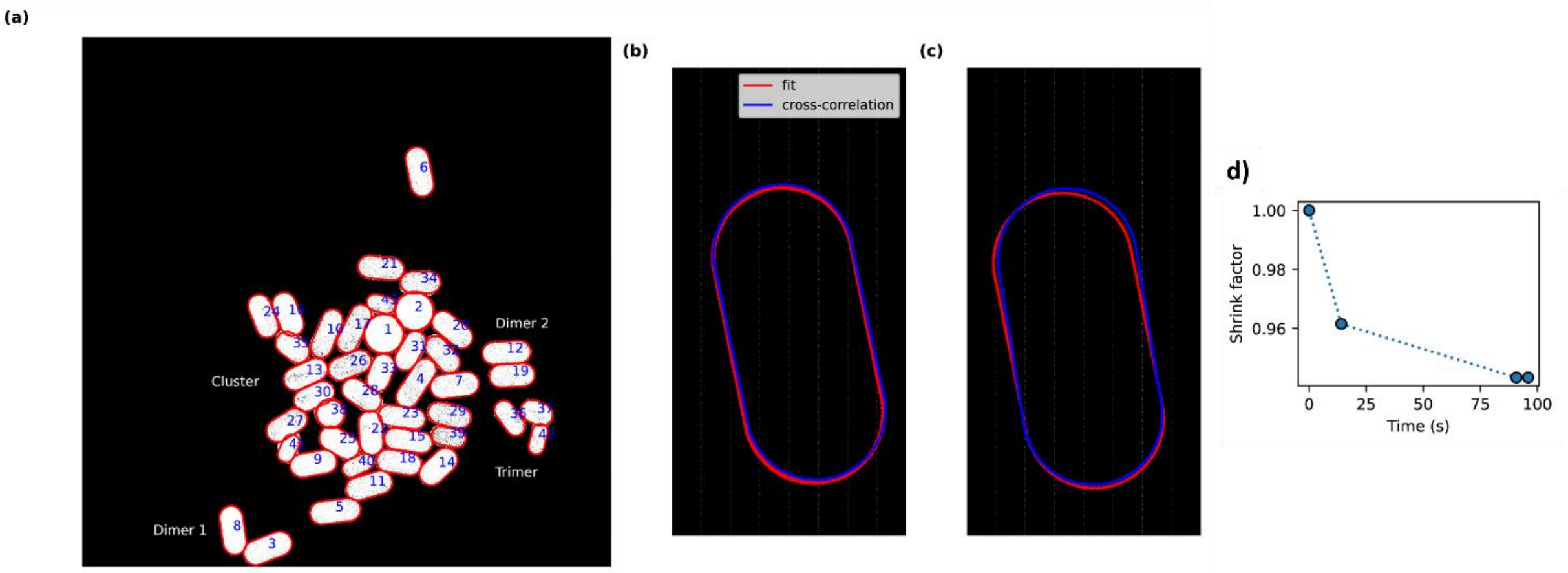


**Fig. S6 Automated segmentation and mask-tracking workflow.** (a) Initial labeling and grouping. (b) Zoom-in of NR 6 after hole filling, showing fitted mask. (c) Mask tracking across frames. (d) Vertical shrinking factor over time.

*NR4 validation set: raw frames, difference maps, and noise boundaries*

NR 4 remains structurally stable throughout the experiment and is used to calibrate dynamic noise boundaries. Supplementary Fig. S7a(i–viii) shows raw TEM frames over the same time interval used in main-text Fig. 3 and Fig. 4. The corresponding difference maps are shown in Supplementary Fig. S7a(ix–xiv), using the color scale introduced in Supplementary Fig. S3.

The full asymmetry trace for NR 4 is shown in Supplementary Fig. S7b, with >95% of values contained within the noise band. The switching rate, shown in Supplementary Fig. S7c, confirms negligible structural dynamics.

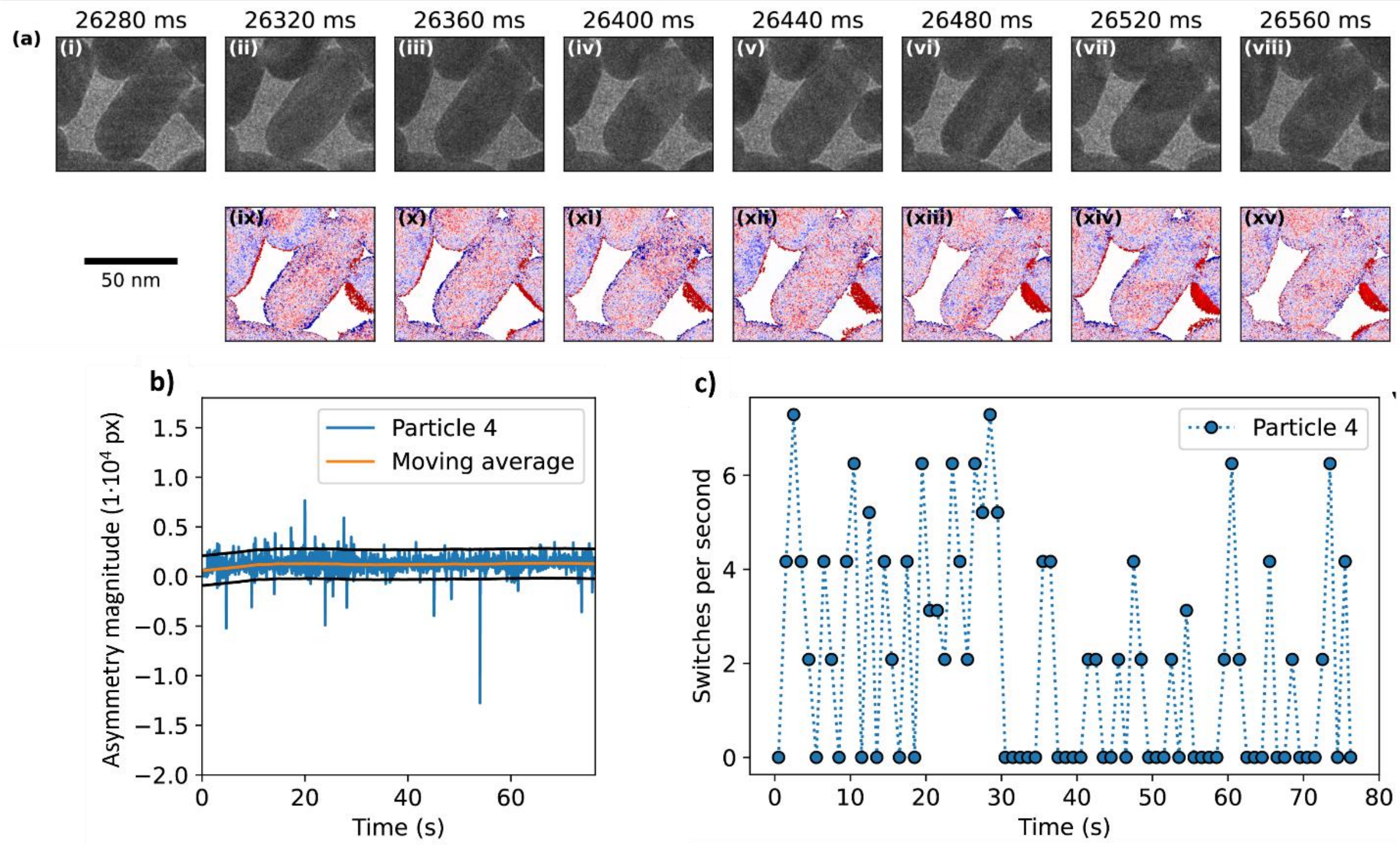


**Fig. S7 Noise calibration using a static reference nanorod (NR 4).** (a) Raw frames (i–viii) and corresponding difference maps (ix–xv). (b) Full asymmetry trace for NR\,4 with noise boundaries. (c) Switching rate confirming structural stability.

*Particle activity*

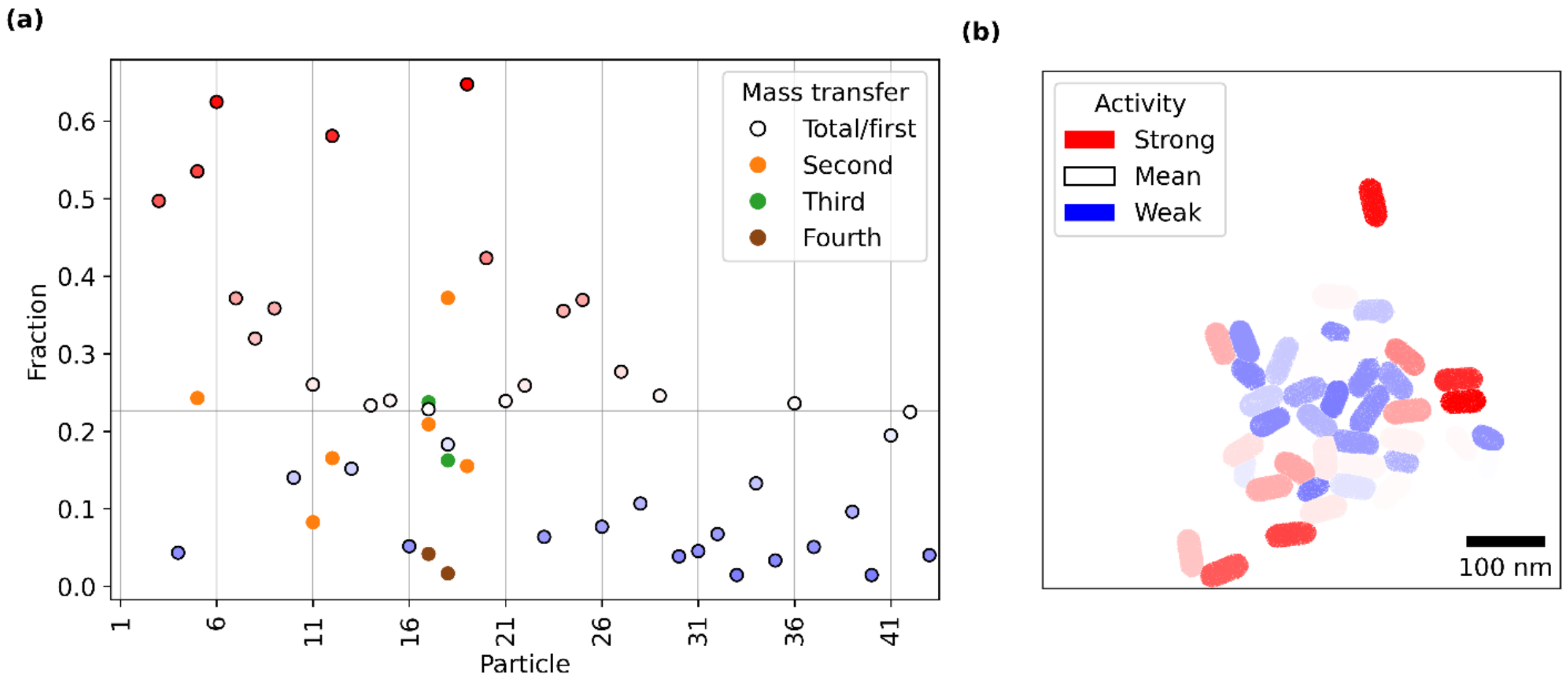


**Fig. S8 Reshaping activity across all particles.** Activity is defined as the fraction of frames in which the tip-asymmetry signal lies outside the noise band calibrated on the static reference rod NR 4. (a) Activity for each particle. For particles undergoing one or more mass-exchange events (as detailed in the discussion surrounding Fig. 4 in the main text), the values before and after the event are shown separately (colours as in the legend); quoted statistics of activity in the main text use the pre-exchange values only, where all particles can be compared, which range from 1.5% to 65% (mean 23% and standard deviation of 17%). The blue-red colour scale before any mass exchange indicate the activity of the particle with red and blue indicating high and low activity, respectively. (b) The spatial map of the activity before any mass exchange shows that the most active particles lie isolated or at the cluster edge while the least active are located within the cluster. The reason is likely that particles inside the clusters are strongly constrained by the increasing connectivity of the cluster.

*Switching rate*

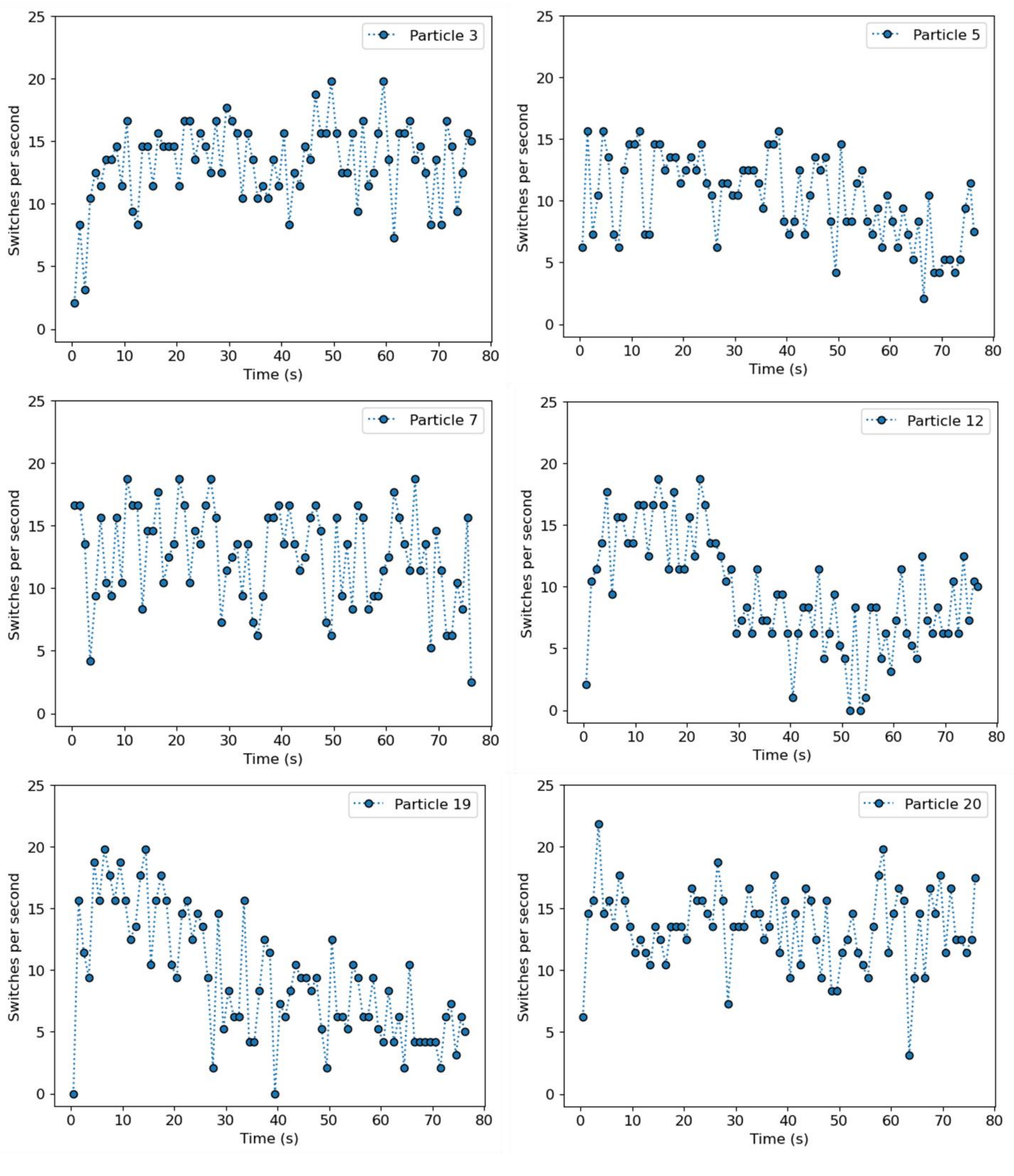


**Fig. S9 Switching-rate traces for six further active particles.** Switching rate over the course of the experiment for six particles selected from the more active part of the ensemble (Fig. S10a), computed as for NR 6 in main-text Fig. 3c. Some particles show a progressive rise comparable to that of NR 6, while others remain at a roughly constant rate throughout; the magnitudes are generally

somewhat lower than for NR 6. The increase in reshaping activity is therefore not observed universally, and its occurrence does not correlate obviously with position within the cluster.

*NR12 and NR19: Full-time asymmetry and switching analysis*

Extended asymmetry analysis for NR 12 and NR 19 is shown in Supplementary Fig. S10. The projected-area traces (Supplementary Fig. S10a,c) reveal the abrupt mass-transfer event and subsequent dynamical arrest. Switching-rate traces (Supplementary Fig. S10b,d) highlight the transition from active reshaping to immobilization after the critical event at t = 26.44 s.

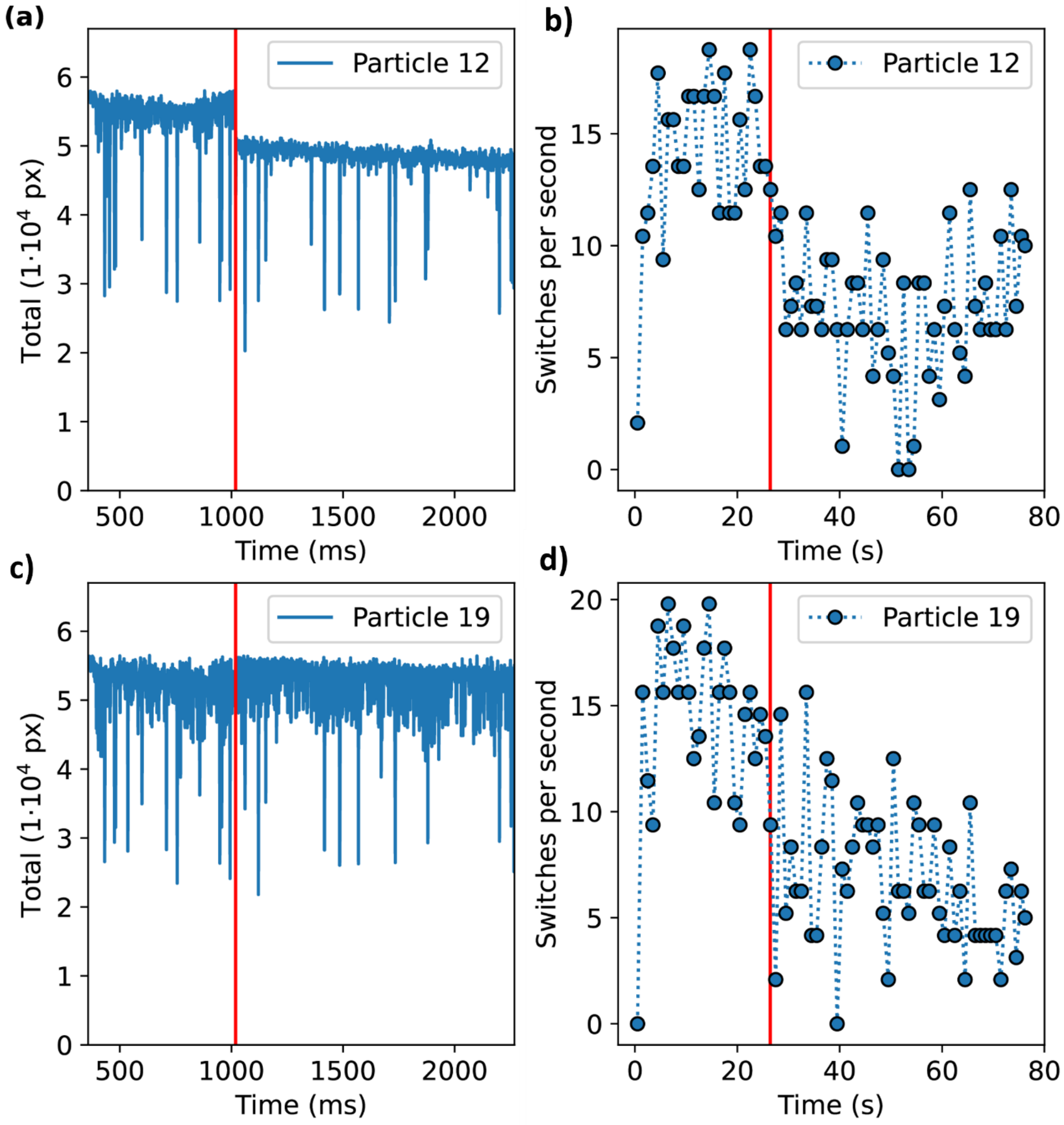


**Fig. S10 Complete asymmetry and switching-rate analysis for NR 12 and NR 19.** (a) NR 12: projected-area trace. (b) NR 12: switching rate. (c) NR19: projected-area trace. (d) NR 19: switching rate.

*Extended dynamic reshaping sequences*

Full difference-map sequences for isolated and coupled reshaping are shown in Supplementary Fig S11. Panel (a) shows an 8-frame window for NR 6 corresponding to the short-interval dynamics in main-text Fig. 3a. Panel (b) shows the corresponding sequence for the coupled pair NR 12 and NR 19, matching the interval in main-text Fig. 4d.

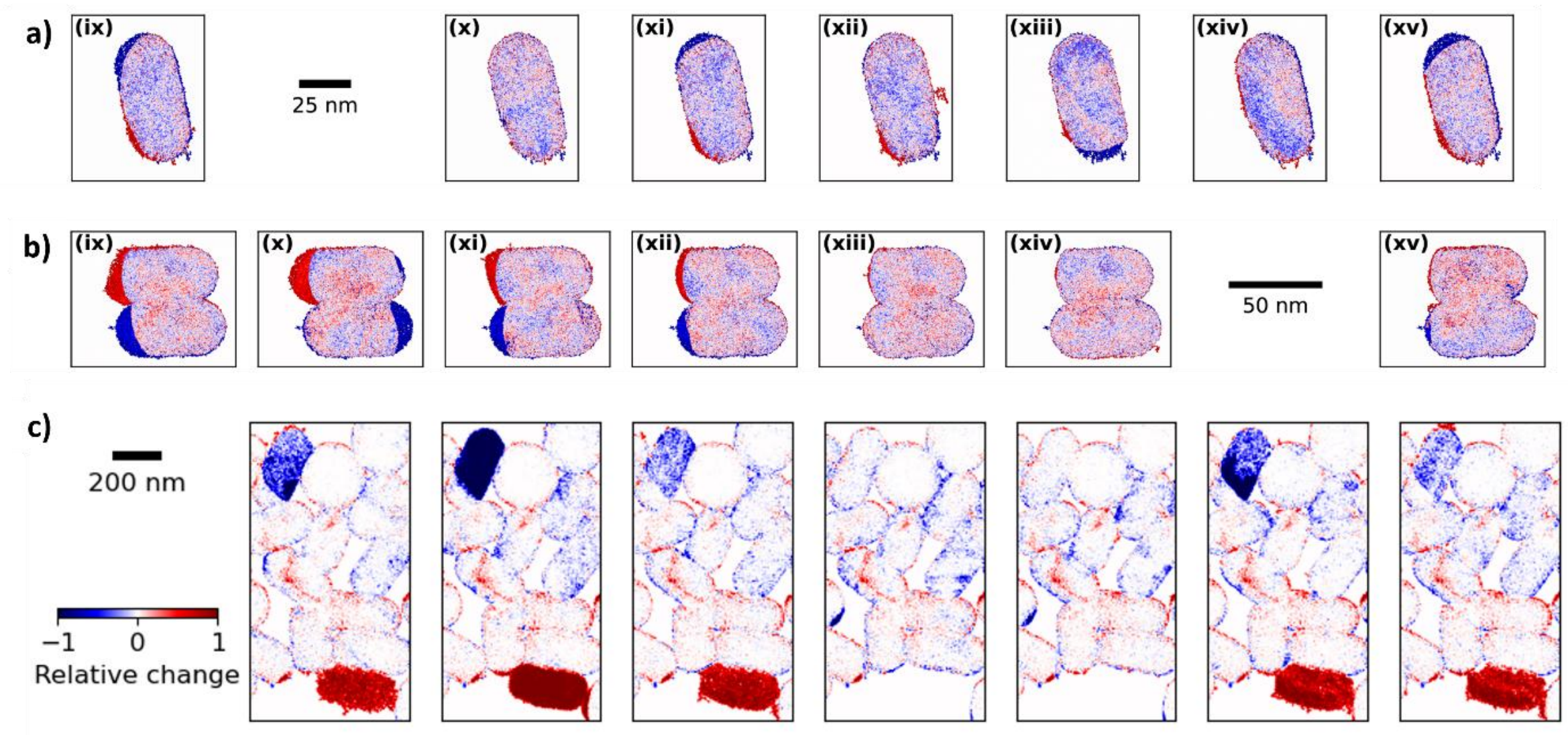


**Fig. S11 Extended dynamic reshaping sequences.** (a) NR 6 difference-map sequence over 8 consecutive frames. (b) NR 12/19 coupled reshaping sequence over the corresponding interval. (c) Zoomed cluster difference-map sequence corresponding to main text Fig. 5a.

*Estimation of surface diffusion lengths per pulse*

To assess whether atomic diffusion can plausibly account for the observed length scales of mass redistribution, we estimated diffusion lengths from a standard random-walk model using literature values for the diffusion coefficient of gold.

The temperature-dependent diffusion coefficient D follows the Arrhenius relation,

$$D_s(T) = D_0 \times e^{-E_a/k_B T}$$

where $D_0$ is the diffusion prefactor, $E_a$ is the activation energy, $k_\beta$ is the Boltzmann constant, and T is the absolute temperature. Atoms diffusing on a surface perform a two-dimensional random walk, but the quantity of interest here is the displacement along the path connecting two particles. We therefore take the one-dimensional root-mean-square displacement as the characteristic diffusion length L over a time t:

$$L = \sqrt{2 \times D_s \times t}$$

We evaluated this at our simulated peak temperature, T = 1115 K (Fig. 1e, Methods), using t equal to either a single optical pulse (400 μs) or a small number of pulses, consistent with the observed timescales of mass redistribution in the main text.

*Bulk diffusion.* For bulk self-diffusion in gold, we used literature activation energy and prefactor values typical of face-centred-cubic metals diffusing via a vacancy mechanism ($E_a \approx 1.7$–$1.8$ eV; $D_0 \approx 10^{-2}$–$1$ $cm^2\ s^{-1}$) [ref]. At 1115 K, this gives a bulk diffusion length of order 1–10 nm per pulse, far shorter than the hundreds-of-nanometre distances over which we observe correlated mass redistribution, and shorter even than the dimensions of a single nanorod.

*Surface diffusion.* For surface self-diffusion, we used literature values for single-adatom hopping on the low-index Au(001) and Au(111) facets: $D_0(001) = 8.0\times10^{-4}$–$1.0\times10^{-2}$ $cm^2\ s^{-1}$ and $E_a(001) = 0.84$ eV; $D_0(111) = 2\times10^{-4}$–$3.7\times10^{-4}$ $cm^2\ s^{-1}$ and $E_a(111) = 0.67$ eV (Jiang et al., Sci. Adv. 10, eadn5946 (2024), and references therein). Table S[X] gives the resulting diffusion coefficient Ds and diffusion length L at T = 1115 K for one pulse (400 μs) spanning the reported range of $D_0$ for each facet.

| Facet | $D_0$ ($cm^2 s^{-1}$) | $E_a$ (eV) | $D_s$ at 1115 K ($cm^2 s^{-1}$) | L, 1 pulse (nm) |
|---|---|---|---|---|
| Au(001), low $D_0$ | $8.0 \times 10^{-4}$ | 0.84 | $1.28 \times 10^{-7}$ | 101 |
| Au(001), high $D_0$ | $1.0 \times 10^{-2}$ | 0.84 | $1.60 \times 10^{-6}$ | 357 |
| Au(111), low $D_0$ | $2.0 \times 10^{-4}$ | 0.67 | $1.87 \times 10^{-7}$ | 122 |
| Au(111), high $D_0$ | $3.7 \times 10^{-4}$ | 0.67 | $3.47 \times 10^{-7}$ | 167 |

**Tab. S1** Estimated surface self-diffusion lengths of gold at T = 1115 K.


\end{document}